\documentclass[lettersize,journal]{IEEEtran}
\usepackage{amsmath,amsfonts}
\usepackage{algorithmic}
\usepackage{algorithm}
\usepackage{array}
\usepackage{textcomp}
\usepackage{stfloats}
\usepackage{url}
\usepackage{verbatim}
\usepackage{graphicx}
\usepackage{cite}
 \usepackage{subfigure}
 \usepackage{siunitx}
\begin{document}

\title{Combining Heuristic and Reinforcement Learning to Achieve the Low-latency and High-throughput Receiver-side Congestion Control}

\author{Xianliang Jiang, Guanghui Gong, and Guang Jin
        % <-this % stops a space
  \thanks{ Xianliang Jiang and Guang Jin are now working at Faculty of Electrical Engineering and Computer Science of Ningbo University, China (e-mail: jiangxianliang@nbu.edu.cn, jinguang@nbu.edu.cn). Guang Jin is the corresponding author.}%
  \thanks{Guanghui Gong is now working at Eastern Institute of Technology, Ningbo, China(e-mail: ggong@eitech.edu.cn).}}% <-this % stops a space}

% The paper headers
\markboth{Journal of \LaTeX\ Class Files,~Vol.~x, No.~x, xx~2024}%
{Shell \MakeLowercase{\textit{et al.}}: A Sample Article Using IEEEtran.cls for IEEE Journals}

\maketitle

\begin{abstract}
Traditional congestion control algorithms struggle to maintain the consistent and satisfactory data transmission performance over time-varying networking condition. Simultaneously, as video traffic becomes dominant, the loose coupling between the DASH framework and TCP congestion control results in the un-matched bandwidth usage, thereby limiting video streaming performance. To address these issues, this paper proposes a receiver-driven congestion control framework named Nuwa. Nuwa deploys the congestion avoidance phase at the receiver-side, utilizing one-way queueing delay detection to monitor network congestion and setting specific target delays for different applications. Experimental results demonstrate that, in most cases, with appropriate parameter configuration, Nuwa can improve the throughput of TCP flows 4\% to 15.4\% and reduce average queueing delay by 6.9\% to 29.4\%. Furthermore, we also introduce the use of reinforcement learning to dynamically adjust Nuwa's key parameter $k$, enhancing Nuwa's adaptability to the unpredictable environment.
\end{abstract}


\section{Introduction}
\IEEEPARstart{T}{HE} Transmission Control Protocol (TCP) is a critical part of the network protocol stack to support reliable data delivery. In recent years, the gradual enrichment of mobile devices, along with the deployment of 5G cellular network architecture has led to a significant increase in the link capacity of the network and a dramatical improvement in wireless network performance\cite{a1:A}. People tend to use mobile devices to connect to wireless networks for interaction than to use wired networks, causing wireless communication traffic to grow rapidly. At the same time, the rise of various applications such as video streaming/meetings, virtual reality(VR), and augmented reality(AR), etc., which typically require high-throughput, low-latency network transmission to ensure a good user experience, has placed higher demands on wireless networks. However, due to the inherent characteristics of wireless networks such as limited bandwidth and susceptibility to interference, even after continuous upgrading and optimization, wireless networks still have shortcomings in processing large amounts of high-speed data transmission, leading to network congestion and reduced throughput, among other issues\cite{a2:V}. In addition, wireless networks greatly enhance user mobility, and changes in user location may affect the strength and quality of signal reception, which in turn will impact the performance and efficiency of wireless network transmission.

Congestion control is an important mechanism in TCPs, which aims to adjust the rate at which sender-side sends data to avoid network congestion and throughput collapse when traffic is overloaded\cite{a3:S}. However, the performance of traditional congestion control algorithms in wireless networks is hardly satisfactory. The main reason is that wireless networks are not as stable as wired networks. Traditional congestion control algorithms have a strong dependency on the network environment, usually based on some assumptions and predictive models to deduce the network state to maintain stable performance\cite{a4:Z}. Due to the influence of numerous interfering factors and base station switching, the wireless network environment is highly variable. On the one hand, this can cause the network capacity to fluctuate greatly, and traditional algorithms may not be able to accurately estimate the available bandwidth and respond to bandwidth changes in a timely manner. On the other hand, the characteristics of the wireless channel make it more prone to packet loss and increased latency, and the mechanism of traditional algorithms is no longer suitable for this situation. 

\begin{figure}
    \centering
    \includegraphics[width=1\linewidth]{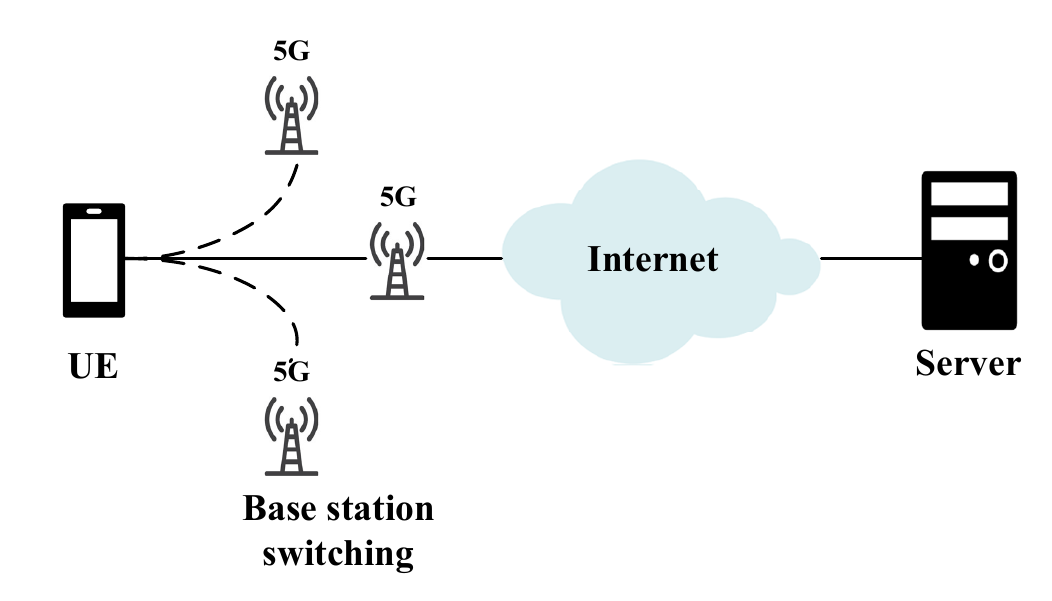}
    \caption{The network topology adopted to verify the performance of traditional congestion control algorithms in 5G networks.
}
    \label{fig1}
\end{figure}

To verify above-mentioned issues, we conduct several file download experiments using the topology shown in Fig.~\ref{fig1} to investigate the transmission performance of traditional congestion control algorithms in 5G mobile networks. The user equipment (UE) is connected to the 5G base station via radio wave signals, the base station network is linked to the WAN through the associated nodes, and the server is deployed in the cloud. In the experiment, the user moves back and forth at a constant speed of approximately \SI{20}{km/h} along a fixed route with six deployed base stations to achieve base station switching. While the user is moving, the UE downloads files from the server. The cloud file is large enough to accommodate a file download of 80 seconds. In the experiment, we utilized China Mobile's 5G commercial network and choose two classic TCP congestion control algorithms, BBR and CUBIC, as the transport layer algorithms. CUBIC\cite{a8:S} is designed for the wired network and can perform well in high latency, high bandwidth network environment, but is generally not suitable for cellular network. BBR\cite{a9:Y} is more suitable for unstable network conditions, such as wireless network and high latency, high packet loss network. The results show that the throughput of CUBIC and BBR exhibits severe fluctuations and often drops to zero. The channel bandwidth is rapidly changing due to multiple factors, and base station switching and user mobility exacerbate this phenomenon.

Through real-time network experiments on the 5G testing platform, \cite{a53:H} points out that CUBIC is likely to result in buffer bloat. Moreover, sporadic high-rate bursts with BBR lead to prolonged RTT when devices move away from access points, thus causing the 5G link to become a bottleneck. \cite{a55:T} investigates the performance of CUBIC and BBR algorithms in a real 5G network environment using 5G devices. It points out that the throughput of these two algorithms exhibits significant fluctuations over the long term. \cite{a56:A} points out that in millimeter-wave links, BBR is not suitable for applications that require both low latency and high throughput simultaneously, as BBR achieves low latency goals by lowering the sending rate. The deployment of 5G networks is gradually expanding, while improving service quality, it also presents new challenges to the performance of classical congestion control algorithms. 

We believe that how to make algorithms quickly capture and respond to changes in network bandwidth is the key to improving wireless network data transmission. Most traditional congestion control algorithms implement congestion control at sender-side, and sender-side needs to understand the network conditions and packet loss through feedback information from the receiving end, making it difficult for sender-side to timely obtain and adjust to changes in the network in a rapidly changing network environment. At the same time, for wireless networks, the last hop may become a bottleneck for data transmission, and the data transmission performance depends on the calculation of the last hop bandwidth. Besides, in recent years, video streaming has seen a significant increase in traffic and gradually taken the lead. DASH(Dynamic Adaptive Streaming over HTTP) is a widely used video streaming protocol, it can select appropriate video chunks for transmission based on the current network condition and bandwidth situation. However, because the DASH protocol is deployed on the receiver-side, it cannot obtain bandwidth information estimated by sender-side. The loose coupling between DASH and TCP results in poor estimation accuracy of channel bandwidth, thus reducing video quality. If congestion control can be implemented on the receiving end, not only can the congestion control algorithm capture bandwidth changes more timely and accurately, but the estimated bandwidth information on the receiving end can also be provided to DASH as reference information to optimize the video transmission quality.

Therefore, we propose Nuwa, a receiver-driven congestion control framework. First, based on the CUBIC algorithm, Nuwa decouples the congestion avoidance phase of sender-side congestion control and implements it at receiver-side. Implementing congestion control at the receiver-side allows the receiver-side to directly perceive the network conditions based on the received packets, thereby more accurately estimating the window size, adapting better to changes in the network environment, and improving transmission efficiency. Secondly, Nuwa modifies the congestion window size by detecting network bandwidth through one-way delay. One-way delay only considers the time it takes for data to be sent from sender-side to receiver-side, enabling a faster and more accurate reflection of network congestion. Finally, Nuwa sets different target delays for each application to meet the varying requirements of different applications for delay.

We conduct extensive experiments in 5G cellular networks, the experimental results show that Nuwa can monitor the bandwidth changes in the wireless network in real-time and adjust the congestion window promptly, effectively improving data transmission performance. However, as a heuristic algorithm, although Nuwa can quickly respond to changes in the network environment, it can only provide the best strategy in a specific network environment, and may not consistently produce the same effect in different network environments. Due to breakthroughs in various new technologies, the network environment is complex and constantly changing, traditional congestion control algorithms are usually developed for specific network environments, algorithm performance often falls short of expectations. To address the above issues, we further integrate Nuwa with reinforcement learning. The agent interacts with the network environment, autonomously learns the characteristics of different network environments, and provides corresponding strategies to adjust Nuwa's key parameters. This enables Nuwa to offer optimal strategies in different network environments to optimize overall performance. It is worth stating that the preliminary results of this paper are discussed at two international conferences. The paper\cite{ab:G} conducts a series of discussions and experiments on the heuristic Nuwa algorithm. Building upon the work presented in\cite{ab:G},
paper\cite{ac:G} proposes the intelligent Nuwa algorithm that combines reinforcement learning. The summary of our contributions is as follows:
\begin{itemize}
    \item We propose a receiver-driven congestion control framework called Nuwa. The Nuwa algorithm implements congestion control at the receiver-side, considering only one-way network transmission state information from the sender-side to the receiver-side to estimate the sending window size and feed it back to the sender-side. 
    \item Nuwa sets different target delays for each application. The setting of the target delay determines Nuwa's ability to occupy bandwidth. Setting a reasonable target delay helps Nuwa obtain fair bandwidth and solves the problem of significant throughput reduction when competing with loss-based algorithms in delay-based methods.
    \item We propose to enhance Nuwa's adaptability in various network environments by incorporating reinforcement learning. By providing performance objectives and utilizing reinforcement learning, we dynamically adjust the key parameter $k$. This enables Nuwa to maintain excellent performance in a broader range of network scenarios.
\end{itemize}

The rest of this paper is organized as follows. Section \uppercase\expandafter{\romannumeral2} discusses related works. Section \uppercase\expandafter{\romannumeral3} explains our motivation. Section \uppercase\expandafter{\romannumeral4} introduces the Nuwa algorithm and provides preliminary performance testing. Section \uppercase\expandafter{\romannumeral5} elaborates on how to use reinforcement learning to further enhance the performance of the Nuwa algorithm. Section \uppercase\expandafter{\romannumeral6} conducts a more comprehensive performance test of the Nuwa algorithm. In Section \uppercase\expandafter{\romannumeral7} we summarize the work of this article and outline future work.

\section{Related Works}
TCP, as a connection-oriented reliable transport protocol, is used by the vast majority of service applications today. Congestion control algorithms are key to ensuring high efficiency of TCP transmission. CUBIC is the default algorithm for the Linux kernel, which treats packet loss as a signal of network congestion. CUBIC uses a cubic function to control the growth of the congestion window, and its congestion window growth is independent of RTT, which allows for good RTT fairness between TCP connections sharing multiple bottleneck links. Unlike CUBIC, delay-based congestion control algorithms treat delay growth as a congestion signal in the network. CUBIC-FIT\cite{a21:J} extends the CUBIC algorithm framework using a delay-based approach. It simulates multiple CUBIC flows in a single TCP connection and adjusts the simulated CUBIC flows using end-to-end queuing delay to improve CUBIC's throughput performance over wireless links. Copa\cite{a23:V} optimizes throughput and delay using a Markov model and detects if the link buffer is full by observing changes in delay. Copa+\cite{a24:W} enhances Copa with a parameter adaptation mechanism and an optimized competitive mode entrance criterion. It can adaptively clear bottleneck buffer occupancy to correctly estimate the base RTT. However, it is too simplistic to judge the network congestion based solely on a single indicator such as packet loss or delay, which may lead to misjudgment of the network conditions. BBR estimates link capacity based on the observed bottleneck bandwidth and round-trip propagation time, and adjusts its data transmission rate to maximize bandwidth utilization without increasing the buffer queue. TCP-Enewreno\cite{a54:M} is designed for 5G network scenarios, where the sender-side estimates the network bandwidth and dynamically adjusts the congestion window based on the estimated bandwidth value. The paper\cite{a57:K} proposes the mmS-TCP algorithm, which modifies the window adjustment mechanism of S-TCP\cite{a58:K}, enhancing the average total throughput and fairness within the protocol. Additionally, by incorporating CoDel\cite{a59:K}, it ensures that the TCP protocol can maintain its performance in millimeter-wave networks. D-TCP\cite{a60:M} estimates the available channel bandwidth, utilizes bandwidth information to derive congestion control factors, and adaptively increases or decreases the congestion window to address fluctuations in millimeter-wave channel conditions. PCC\cite{a37:M} points out that even with extensive modifications, it is difficult to achieve sustained high performance for congestion control algorithms based on TCP's congestion control framework. Therefore, PCC proposes a performance-oriented congestion control framework where the sender-side adjusts its transmission strategy based on observed performance metrics.

The congestion control algorithm mentioned above, based on the sender-side, can only perceive the network condition indirectly. The sender-side adjusts the sending window based on the ACK feedback sent by the receiver-side to understand the degree of network congestion. In wireless networks, especially in high-speed networks like 5G, the sender-side faces difficulty in accurately and promptly perceiving the congestion status of the network. In comparison to sender-side, receiver-side can directly perceive the network and obtain real feedback from the network. Therefore, there are now many congestion control schemes that address issues with TCP transmission by implementing improvements at receiver-side. In DRWA\cite{a25:H}, a countermeasure implemented at receiver-side is designed to solve the buffer problem in resource-constrained environments such as Wifi. Receiver-side increases its receive window to make the RTT closer to its minimum RTT. In DFCSD\cite{a26:P}, AQM (Active Queue Management) is integrated into a loss-based congestion algorithm by controlling the RTAC\cite{a27:H} of the cellular network and implemented at the receiver-side to meet application requirements for high throughput and low latency. HRCC\cite{a37:B} proposes a hybrid receiver-side congestion control framework for WebRTC (Web Real-time Communication). The framework leverages reinforcement learning to observe network links and periodically adjust the estimated bandwidth using heuristic schemes. RBBR\cite{a38:H} present a receiver-driven BBR algorithm, which differs from the original BBR algorithm by estimating the sending rate at receiver-side. The paper\cite{a64:Y} proposes a receiver-side traffic control algorithm that adjusts the sender's upstream buffer by monitoring available upload capacity and dynamically adjusting the receive window. Homa\cite{a39:B} is a receiver-driven low-latency transport protocol for data center scenarios. It uses priority queues in the network to ensure low latency for short messages, and receiver-side dynamically manages priority allocation. AMRT\cite{a40:J} proposes a receiver-driven transmission scheme that uses anti-ECN(Explicit Congestion Notification) marking to increase the transmission rate in situations where link utilization is insufficient. RPO\cite{a41:J} retains the advantage of receiver-driven transmission while making reasonable use of low-priority opportunity packets and ECN marking to improve network utilization.

With the diversification of network scenarios and the complexity of network environments, manual summarization of different network scenario characteristics and the targeted design of corresponding heuristic congestion control algorithms are no longer sufficient to solve the problems of network transmission. Therefore, learning-based congestion control algorithms have begun to receive more and more attention, as RL agents can provide optimal decisions through interaction with the network environment and autonomous learning to further optimize network transmission. Remy\cite{a61:K} is an early attempt to model congestion control, treating it as a partially observable Markov decision process problem, adjusting the transmission window differently for various network states. Aurora\cite{a28:N} is the first to use deep reinforcement learning to drive an agent to automatically explore network congestion control strategies, mapping observed network statistics (such as latency and throughput) to rate selection. Eagle\cite{a62:S} is an online learning algorithm that combines expert knowledge with deep reinforcement learning, enabling the algorithm to adapt to new network conditions. QTCP\cite{a63:W} combines TCP with Q-learning algorithm, allowing the sender-side to learn the optimal congestion control strategy online. Orca\cite{a29:S} combines classical congestion control strategies with advanced modern deep reinforcement learning (DRL) techniques for congestion control, using CUBIC as the underlying logic and deep reinforcement learning for coarse-grained regulation. AUTO\cite{a30:X} proposes a multi-objective reinforcement learning-based adaptive congestion control method, where the policy agent can generate optimal policies for all possible network states and preferences. Libra\cite{a31:Z} proposes a unified congestion control framework that enhances flexibility, adaptability, and practicality by combining the wisdom of classical and reinforcement learning-based congestion control algorithms. RayNet\cite{a32:L} points out that learning-based congestion control protocols are still in the early stages, and proposes a scalable and adaptable simulation framework for developing learning-based network protocols to promote the development of learning-based network protocols. 

Additionally, the diversification of applications has posed challenges for network transmission. Compared to other applications, video streaming requires larger bandwidth and lower latency to ensure a satisfactory user experience. For video streaming, client-side video players typically use Adaptive Bitrate (ABR) algorithms to optimize Quality of Experience (QoE) for users. Pensieve\cite{a42:H} proposes an ABR algorithm based on reinforcement learning, which trains a neural network based on observation results collected by the client-side video player to select appropriate bitrates and resolutions for future video chunks. ABRaider\cite{a43:W} proposes a multi-stage reinforcement learning consisting of offline and online phases. In the offline phase, ABRaider integrates the advantages of various ABR algorithms and formulates strategies suitable for different environments, while in the online phase, it focuses on learning the individual user's environment. However, to improve the quality of video services, it is necessary to consider how to reduce end-to-end congestion delay. The GCC\cite{a44:G} algorithm proposes for video conferencing applications mainly consists of sender-based congestion control based on packet loss and receiver-based congestion control based on delay. Sender-side synthesizes the results of both control algorithms to obtain a final sending rate. Iris\cite{a45:T} utilizes a congestion control model based on statistical learning to control the sending rate and enhances the algorithm's adaptability to the environment by updating the adjustment step size through online learning. Reference\cite{a46:A} explores how to achieve optimal streaming in a 5G environment and provides guidelines for implementing it in 5G networks using the Self-Clocked Rate Adaptation for Multimedia (SCReAM)\cite{a47:I} congestion control algorithm as an example.

To maintain optimal network data transmission performance, congestion control algorithms should accurately identify changes in network bandwidth and rapidly and appropriately respond to these changes. Additionally, the adaptability of congestion control algorithms to the application is crucial. When designing congestion control algorithms, the characteristics and requirements of the upper-layer applications should be taken into consideration.

\section{Motivation}

 \subsection{Loose Coupling between DASH and TCP in Video Streaming}
 DASH is a video streaming technology that selects the appropriate bitrate and resolution to achieve a better viewing experience based on current network conditions. Therefore, accurate bandwidth prediction is crucial for DASH, especially in highly dynamic mobile network scenarios. It can help DASH avoid issues such as reduced video quality or buffering caused by excessively high or low bitrates selected.

\begin{figure*}[!t]
\centering
\subfigure[Fixed Bandwidth]{
\includegraphics[width=6cm]{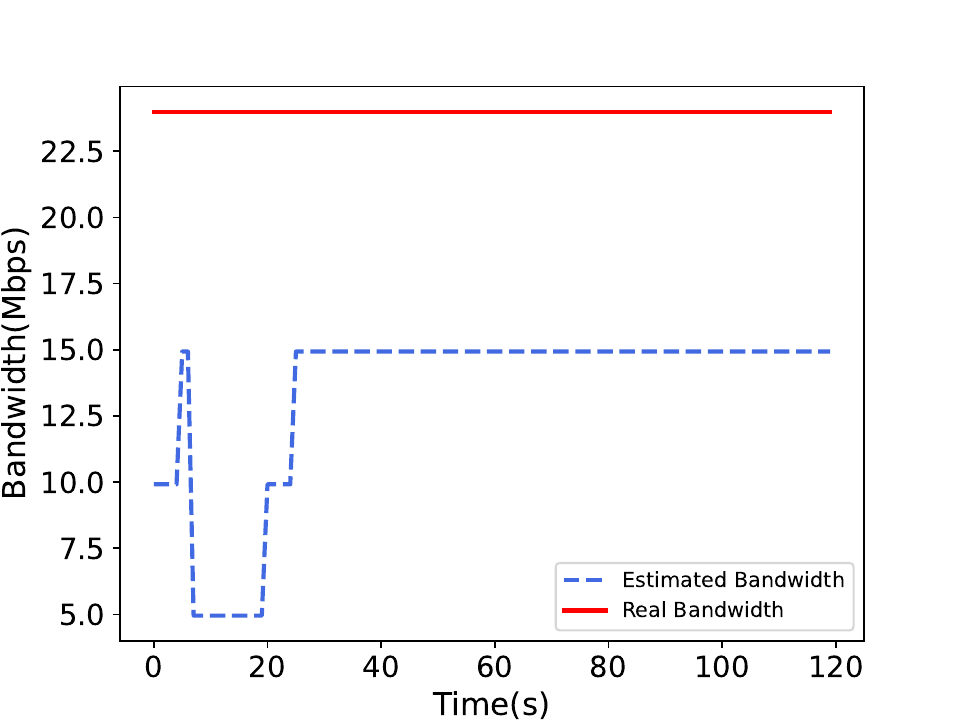}
}%
\subfigure[4G]{
\includegraphics[width=6cm]{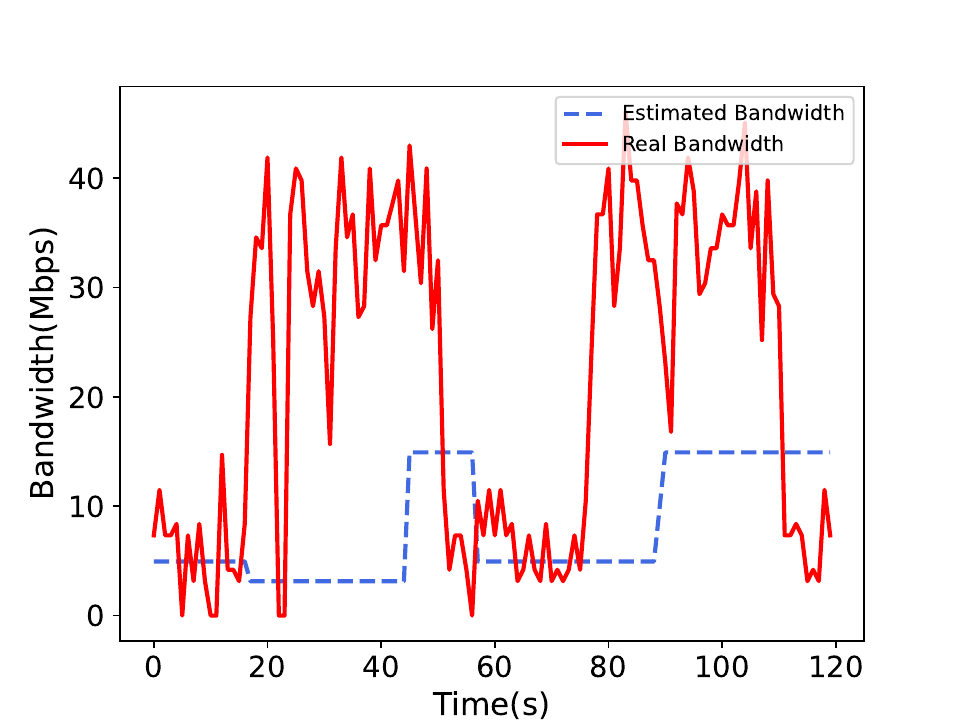}
}%
\subfigure[5G]{
\includegraphics[width=6cm]{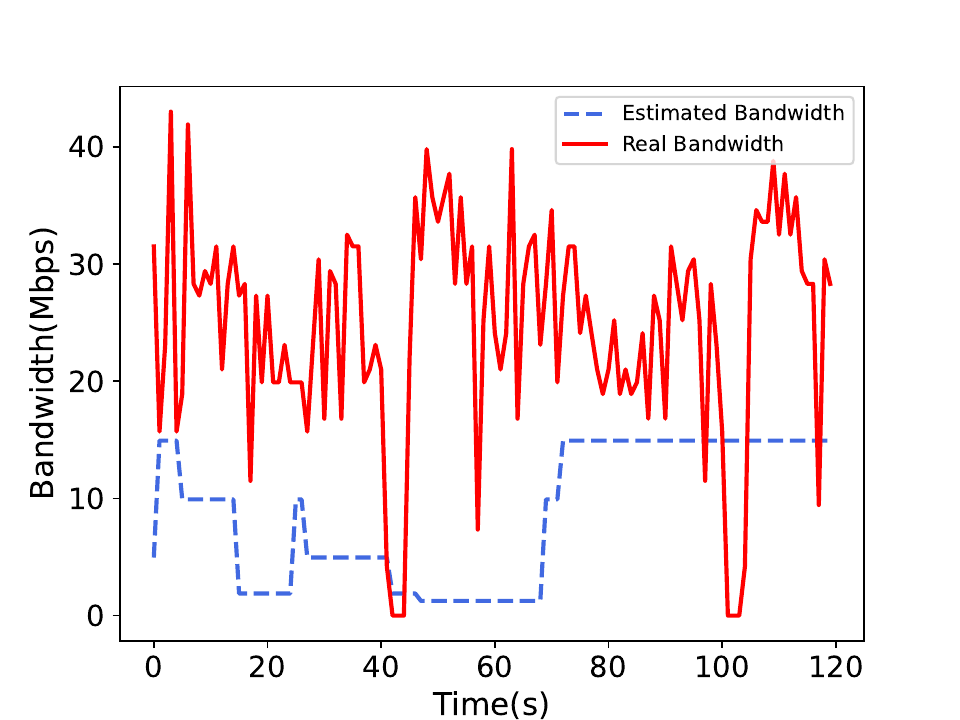}
}%
\caption{Comparison of DASH predicted bandwidth with real bandwidth.}
\label{figg}
\end{figure*}

DASH typically uses its own bandwidth estimation method to predict network bandwidth, such as by calculating the download time and size of each video frame to estimate the current network bandwidth. However, through experiments, we find that there is a discrepancy between the bandwidth values estimated by DASH and the actual bandwidth values. Using the topology structure shown in Fig.~\ref{fig5}, we conduct video streaming experiments in fixed bandwidth, 4G, and 5G network scenarios to test the difference between the bandwidth values estimated by DASH and the actual bandwidth values. In the fixed bandwidth scenario, the bandwidth value is set to 24Mbps, while 4G and 5G use real bandwidth traces we collect. We run the video stream at the receiving end and collect its estimated bandwidth value. Each experiment last for 120 seconds, and the results are shown in Fig.~\ref{figg}. It can be seen that in the fixed bandwidth network scenario, the bandwidth value estimated by DASH is always much lower than the actual bandwidth value. In the 4G and 5G network scenarios, the bandwidth value estimated by DASH is generally much lower than the actual bandwidth value, and in a few cases, it may be higher than the actual bandwidth. Clearly, the accuracy of DASH's bandwidth estimation is poor, which can affect its transmission performance.

We believe that the loose coupling between DASH and TCP is one of the reasons for the inaccurate bandwidth estimation of DASH. In the TCP protocol, congestion control algorithms are typically implemented at sender-side. Sender-side estimates the available network bandwidth and adjusts the size of the sending window and the rate at which data is sent based on feedback information from the network, such as received ACK packets and retransmissions due to timeouts. The size of the sending window is an indicator of the current network bandwidth, which is more closely related to the true bandwidth. However, because DASH is deployed at the receiving end, it cannot obtain the relevant information about the available bandwidth estimated by TCP at sender-side. Therefore, we propose that if a congestion control framework driven by the receiving end could be implemented to estimate the available window size at sender-side, then DASH located at the receiving end could obtain this information and use it as a reference for bandwidth estimation, thereby improving the accuracy of its own bandwidth estimation.
 
 \subsection{Limitations of Sender-side Congestion Control}
 Currently, achieving high throughput and low latency for data transmission in wireless networks is a pressing issue. Enhanced Mobile Broadband (eMBB), as one of the 5G application scenarios, supports high-speed data transmission while providing lower latency, which to some extent improves the performance of wireless network transmission to meet the needs of new applications\cite{a5:M}. However, compared with wired networks, the transmission environment of wireless networks is more complex, and the limited coverage range of wireless networks also restricts their transmission performance. Especially in mobile scenarios, to maintain network connection, mobile devices need to regularly detect the signal strength of surrounding base stations and decide whether to switch base stations based on certain switching strategies, which causes high network capacity fluctuations\cite{a6:M}. In traditional TCP protocol, congestion control is implemented at sender-side. However, sender-side cannot directly perceive the degree of network congestion and can only infer the congestion situation through feedback information transmitted by receiver-side, such as packet loss and delay, which are often affected by network latency and packet loss during transmission. This indirect way of perceiving network congestion may lead to delayed or inaccurate congestion control responses. In contrast, receiver-side can directly observe the network feedback and dynamically adjust the window size based on the actual network conditions, in order to better adapt to network changes. Additionally, receiver-side can purely observe the one-way delay during the data transmission process from sender-side to receiver-side, which more directly reflects the impact of network congestion on data transmission time.

In addition, to implement congestion control at the sender-side, sender-side needs to consider two factors: 1) the available bandwidth of the network, which is the rate at which sender-side sends data; 2) the amount of data receiver-side can accept, which is the size of the receiving window. Therefore, sender-side needs to calculate the congestion window and receiving window based on these two factors and the minimum value of the two is taken. The sending window is constrained by the receiving window and congestion window. In the past, the receiver-side did not have enough time to process the data and cause its buffer to overflow, resulting in problems such as packet loss and retransmission, and even exacerbate network congestion. With the continuous upgrading of mobile device hardware and the development of wireless network technology, the receiving buffer space is no longer a bottleneck for data transmission, and the buffer at receiver-side is adjusted dynamically by Dynamic Receiving Buffer Size (DRS)\cite{a10:W}. However, in some cases, the receiving window may be limited by the congestion window. Increasing the receiving window may cause the congestion window to grow too fast, leading to network congestion. Therefore, DRS increases the size of the receiving window only in cases where the growth of the congestion window is not limited, in order to avoid network congestion\cite{a11:H}. This may result in lower space utilization of the receiving buffer. By implementing congestion control at receiver-side, receiver-side only needs to calculate the available window size for the next data transmission based on the actual network feedback, and transmit the window value to sender-side. This can effectively solve the drawbacks of the congestion mechanism based on congestion window and receiving window, and the window adjustment is more in line with network requirements.

\section{The Proposed Nuwa Algorithm}
In this section, we propose the Nuwa algorithm, elaborate on its design scheme, and discuss its feasibility.
 
\subsection{Overview of Nuwa}

\begin{figure}
    \centering
    \includegraphics[width=1\linewidth]{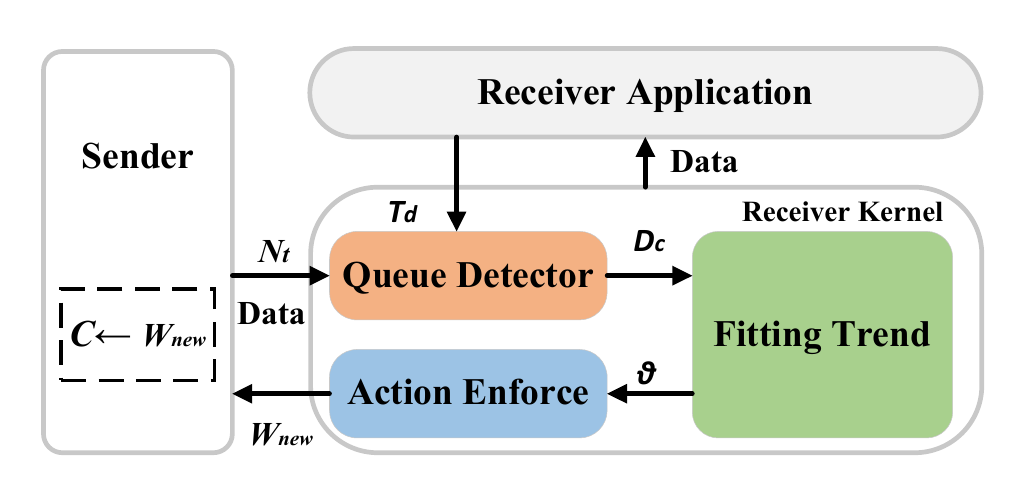}
    \caption{The overall architecture of Nuwa.
}
    \label{fig3}
\end{figure}

Based on the motivation analysis of Section \uppercase\expandafter{\romannumeral3} , we propose a receiver-driven congestion control framework named Nuwa. We decouple the congestion avoidance phase of sender-side congestion control and implement it at receiver-side, so that Nuwa only controls the link during the congestion avoidance phase and the window value is estimated by receiver-side. The design architecture of Nuwa is shown in Fig.~\ref{fig3}, where $\theta $ represents the parameter indicating the trend of window changes, $C$ represents the sending window, and $w_{new}$ represents a new window computed by the receiver-side. $N_{t}$ is the statistical vector of network data sending status. Nuwa consists of two main parts: receiver-side part and sender-side part. Receiver-side part mainly includes Queue Detector module, Fitting Trend module, and Action Enforce module. Queue Detector module is used to estimate queuing delay. Fitting Trend module is employed to observe trends in window changes. Action Enforce module is utilized to calculate the size of the sending window. Receiver-side provides the updated value of the window based on the actual network conditions, which is then transmitted to sender-side as an estimated value of the sending window size. Receiver-side part of the algorithm can be replaced by a kernel dynamic loading module. 

The new window value, $w_{new}$, is conveyed to the sender-side through the Window Size field in the header of ACK packets. With this information, the sender-side can determine the current capacity of the receiver-side to accept data. The idea of determining the sending window by the sender-side is to take the minimum value between the returned receiving window of receiver-side and the estimated congestion window of sender-side. We utilize this mechanism so that, after being returned to the sender-side via ACK, $w_{new}$ will also be assigned to the congestion window, replacing the originally congestion window size estimated by the sender-side. Consequently, the sender-side will ultimately send data with $w_{new}$ as the size of the sending window. In short, $w_{new}$ not only represents the data reception capability of the receiver-side but also indicates the network's congestion situation.

\subsection{Nuwa Algorithm}

The Queue Detector is used to estimate the queuing delay in the wireless network based on the information in the vector $N_{t}$. Nuwa calculates the one-way queue delay of the link to infer the level of congestion in the link. Compared with round-trip delay, one-way delay can more accurately reflect the congestion situation of the network, which helps Nuwa better understand the network conditions and make corresponding adjustments. $N_{t} =\left ( Q_{t}, Q_{m}, R_{t}, R_{m}\right )$ and we use Eq.~\ref{eq1} to calculate the one-way queuing delay:

\begin{equation}
D_{c} =\left ( Q_{t} - Q_{m}\right ) -\left (  R_{t} - R_{m} \right ) \label{eq1} 
\end{equation}

The $Q_{t}$ is the send time of the current packet, $R_{t}$ is the receive time of the current packet, $Q_{m}$ and $R_{m}$ are the send time and receive time of the packet with the smallest one-way delay recorded. Clock synchronization is not required to calculate the delay using the above equation. $D_{c}$ represents the measured one-way queuing delay, whose value indicates the level of congestion in the network. The larger the value of $D_{c}$, the more severe the congestion in the network. In reality, there is a certain discrepancy between the measured value $D_{c}$ and the true value $Q_{d}$, as $D_{c}$ also includes differences in transmission time. We refer to the paper by GCC\cite{a44:G} and use the Kalman filter\cite{a52:R} for data processing. We provide a model for the measured one-way queuing delay, $D_{c}$, as shown in Eq.~\ref{eqa}. $\bigtriangleup M\left ( t \right )$ represents the difference in the amount of data transmitted between the current group and the group in the sliding window with the minimum measured one-way delay, $B\left ( t \right )$ represents the bottleneck link capacity, the ratio of the two indicates the variation in transmission time. $\sigma \left ( t \right )$ is the measurement noise. $Q_{d}\left ( t \right )$ is the one-way queuing delay we wish to obtain after filtering out the transmission time differences and measurement noise.

\begin{eqnarray}  \label{eqa} 
D_{c} \left ( t \right )&=&\frac{\bigtriangleup M\left ( t \right ) }{B\left ( t \right ) } +  Q_{d}\left ( t \right ) +\sigma \left ( t \right ) \nonumber \\
~&=&H\left ( t \right )\cdot \eta \left ( t \right ) + \sigma \left ( t \right )  
\end{eqnarray}

\begin{equation}
\eta \left ( t+1 \right ) =\eta \left ( t \right ) + \phi \left ( t \right )  \label{eqb} 
\end{equation}

After rewriting, $\eta \left ( t \right ) =\begin{bmatrix}\frac{1}{B\left ( t \right ) }  & Q_{d}\left ( t \right ) \end{bmatrix}^{\top }$represents the system state vector,  $H\left ( t \right )=\begin{bmatrix}\bigtriangleup M\left ( t \right ) & 1\end{bmatrix}$. Eq.~\ref{eqb} provides the system state model, $\phi \left ( t \right )$ is the state noise. Based on the system state model and the measurement value $D_{c}$, we can use a Kalman filter to estimate the system's state $\eta \left ( t \right )$. By utilizing the Kalman gain $K_{g}$ to apply the following correction to the estimation, the optimal estimate of the one-way queueing delay can be obtained:

\begin{eqnarray}  \label{eqc} 
Q_{d}\left ( t \right )=&(1-K_{g}\left ( t \right ) )\cdot Q_{d}\left ( t-1 \right )+ \nonumber\\
&K_{g}\left ( t \right )\cdot \left ( D_{c}\left ( t \right )- \frac{\bigtriangleup M\left (t  \right ) }{B\left ( t-1 \right ) }  \right ) 
\end{eqnarray}

We also set different target delays $T_{d}$ for different applications at receiver-side to meet different application requirements and achieve better data transmission. Therefore, $T_{d}-Q_{d}$ represents the direction and magnitude of the window change and adjustment. Its value being positive indicates that the window can be appropriately increased, while its negative value indicates the need to reduce the window, with the magnitude of adjustment increasing as the absolute value increases.

\begin{figure}
    \centering
    \includegraphics[width=0.95\linewidth]{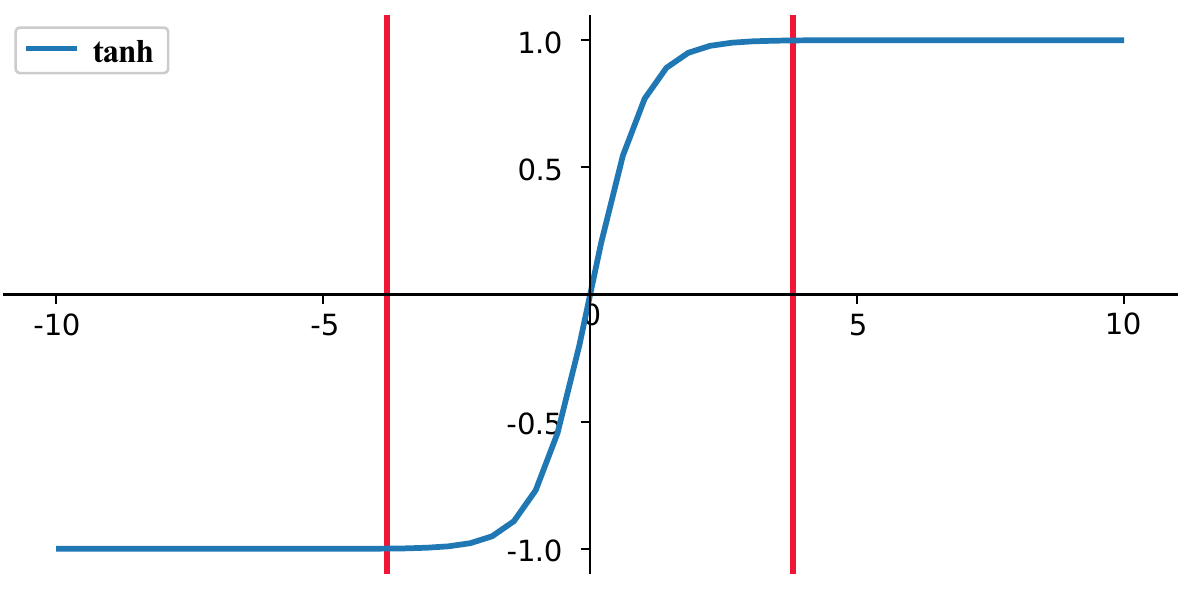}
    \caption{Trend of tanh function.
}
    \label{fig4}
\end{figure}

After detecting changes in the network, Nuwa uses Fitting Trend to calculate the trend of window changes. In order to improve the algorithm's flexibility in response to network changes, we use the {\tt{tanh}} function to calculate the trend of the window. Since Linux kernel does not support floating-point operations, we implement the {\tt{tanh}} function in the kernel using Table calculus. The calculation formula of {\tt{tanh(x)}} is shown in Eq.~\ref{eq2}. Here, we use $x=\frac{T_{d}-Q_{d}}{\rho }$ as the input to {\tt{tanh}}, where $\rho $ is the sensitivity coefficient of queue delay fluctuations. The trend of {\tt{tanh}} is shown in Fig.~\ref{fig4}. The reason for choosing the {\tt{tanh}} function is twofold. On one hand, the function's values must have both positive and negative values to facilitate enlarging or reducing the window. On the other hand, as $x$ approaches zero, we want the function to remain sensitive to network changes. When $x$ approaches zero, the function's derivative is large, indicating rapid changes in the vicinity of that point. Thus, even with small changes in $x$, the function's values will vary to a certain degree. We denote the value of {\tt{tanh(x)}} as $\theta$.

\begin{equation}
tanh\left ( x \right ) =\left (e^{x}-e^{-x}\right ) / \left (e^{x}+e^{-x}\right ) \label{eq2} 
\end{equation}

Finally, we determine the size of the window in the Action Enforce module, and the window update formula is shown in Eq.~\ref{eq3}:

\begin{equation}
w_{new}\gets w_{old}+\left ( \theta \cdot k \right ) / w_{old} \label{eq3} 
\end{equation}

Where $k$ is the aggressiveness parameter for window adjustment, used to control the magnitude of window adjustment. $w_{old}$ represents the previous window size. We use $w_{old}$ as one of the parameters in the window update formula so that the window does not change too much in a single adjustment.

\begin{algorithm}
	\caption{Heuristic Nuwa Algorithm}
	\label{alg1}
	\begin{algorithmic}[1] 
		\STATE Initial:
        \STATE $T_{d}\gets0, Q_{m}\gets500\left ( ms \right ), R_{m}\gets0 $;
        \STATE $Delay_{min}\gets a\; large\; value$;
        \STATE $\rho\gets a\;suitable\;value$;
		\FOR{each ACK} 
        \STATE $Q_{t}\gets Timestamp \; of\; packet\; delivery$;
        \STATE $R_{t}\gets Timestamp\; of\; packet\; receive$;
        \STATE $D_{c}= \left ( Q_{t}-Q_{m}\right )-\left ( R_{t}-R_{m}\right )$;
         \STATE $Q_{d}\gets Filtering \; D_{c}\; with\; Kalman\;Filter$;
		\IF {$Delay_{min}=0 \parallel Q_{d} < Delay_{min} $}
        \STATE $Q_{m}\gets Q_{t}$;
        \STATE $R_{m}\gets R_{t}$;
        \STATE $Delay_{min}\gets Q_{d}$;
		\ENDIF
        \STATE $x\gets\frac{T_{d}-Q_{d}}{\rho }$;
        \STATE $\theta \gets tanh \left ( x\right )$;
        \STATE $w_{new}\gets w_{old}+\left ( \theta \cdot k \right ) / w_{old}$; 
		\ENDFOR
	\end{algorithmic}
\end{algorithm}

The details of the Nuwa algorithm are described in Algorithm~\ref{alg1}. For each ACK, we record the timestamp information of the sending and receiving times of the currently received packet (Lines 6-7). Additionally, we also record the sending and receiving times of the packet with the minimum time difference and update the minimum delay (Lines 10-13). By default, the minimum time difference indicates that there is no queuing delay in the current link or there is a slight queuing delay. Finally, we estimate the one-way queuing delay of the link using Eq.~\ref{eq1}and Eq.~\ref{eqc} (Line 8-9). When $Q_{d}$ is known, we use $T_{d}-Q_{d}$ to indicate the direction and magnitude of the adjustment that the current network should take (Line 15), and use Eq.~\ref{eq3} to update the window (Lines 16-17).

Nuwa is based on the traditional TCP architecture, and only requires slight modifications to implement traditional congestion control algorithms. We only improve the congestion control algorithm during the congestion avoidance phase, which can effectively avoid the typical throughput degradation during data transmission. Based on the idea of one-way delay, Nuwa can more accurately estimate queue delay, thereby understanding the more realistic network conditions, and the receiver-driven approach enables Nuwa to make adjustments more quickly. In addition, Nuwa can operate independently of other algorithms in wireless networks. In Wifi or other wireless network environments, as long as $T_{d}$ is set reasonably, Nuwa can maintain good bandwidth allocation fairness with other algorithms in the network environment.

\subsection{Preliminary Evaluation}
In this section, we conduct preliminary tests on the performance of Nuwa and use CUBIC as a comparative algorithm. A more detailed evaluation of Nuwa's performance is presented in Section~\ref{Sec}.

\subsubsection{Experimental Settings}

\begin{figure}
    \centering
    \includegraphics[width=0.95\linewidth]{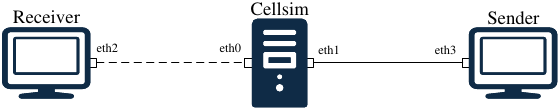}
    \caption{Topology of the emulation experiment of Nuwa.
}
    \label{fig5}
\end{figure}

\begin{figure}
    \centering
    \setlength{\abovecaptionskip}{-0.1cm}
    \includegraphics[width=1\linewidth]{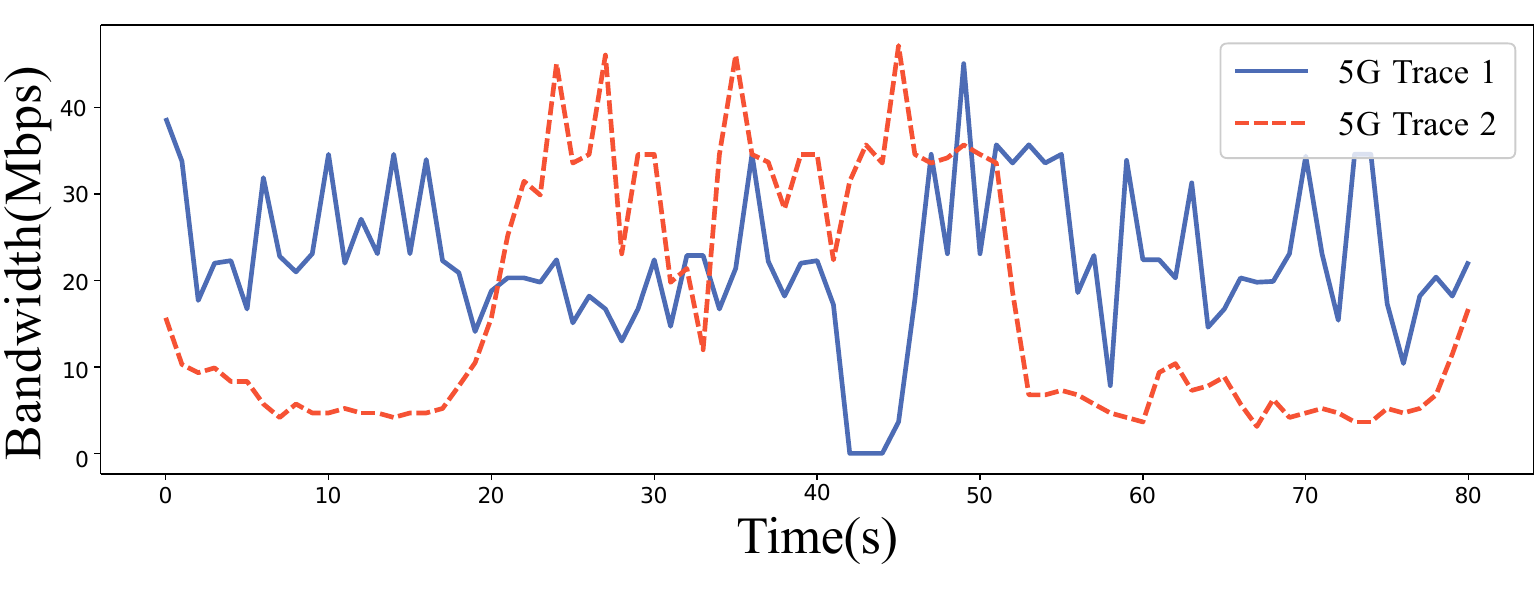}
    \caption{5G traces used in emulation experiments.
}
    \label{fig6}
\end{figure}

In the emulation environment, we conduct emulation experiments using the topology shown in Fig.~\ref{fig5}. We build an experimental platform using three computers, with data being transmitted from the sender-side to the receiver-side using the iperf tool to generate traffic. In the middle of the link, we use a computer as the operating environment for Cellsim\cite{a51:K}, which is a network simulator used for traces replay. In Cellsim, we utilize traces collected in a real network environment as the experimental environment.  Fig.~\ref{fig6} shows two 5G network traces we collect.

\subsubsection{Robustness}

We test the robustness of the Nuwa algorithm to network fluctuation using the experimental platform described above. We choose 5G Trace2 as the simulated experimental network environment, as shown in Fig.~\ref{fig6}, this bandwidth trace has significant fluctuations, with two relatively large network bandwidth changes at around 20 seconds and 55 seconds, which can better demonstrate whether Nuwa can respond well to network fluctuations. We evaluate the robustness of the algorithm to network fluctuation from four perspectives: RTT, throughput, sending window changes, and queue delay.

\begin{figure*}[!t]
\centering
\subfigure[Cwnd Size]{
\includegraphics[width=4.35cm]{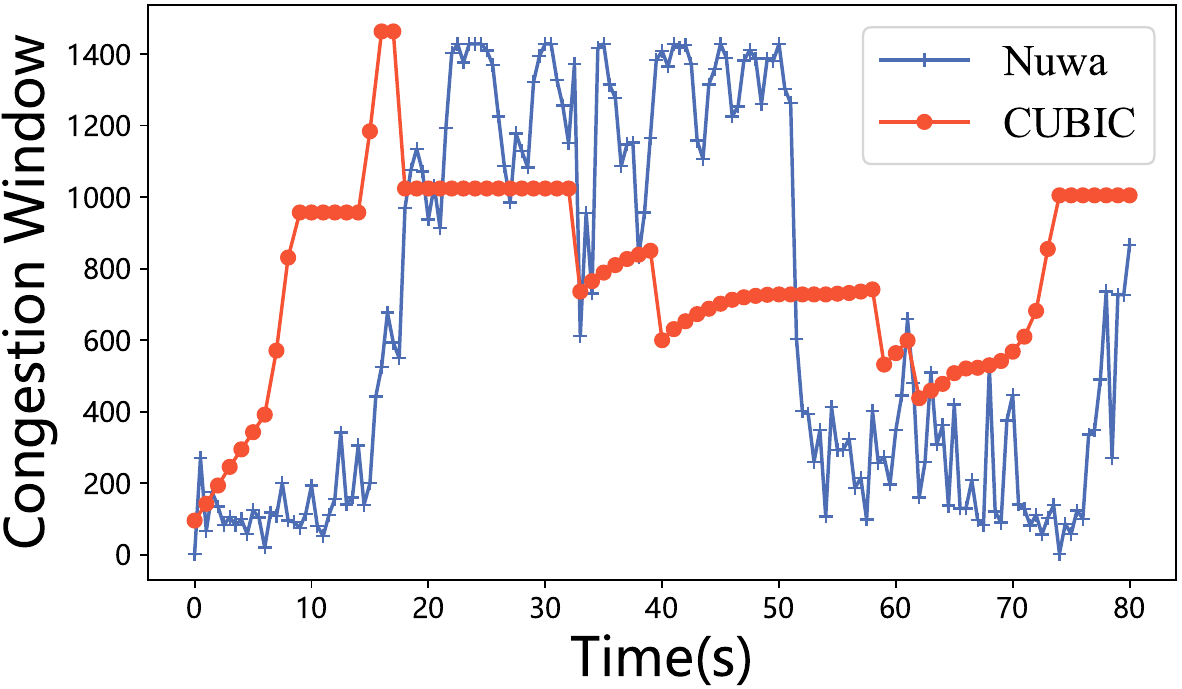}
}%
\subfigure[RTT]{
\includegraphics[width=4.35cm]{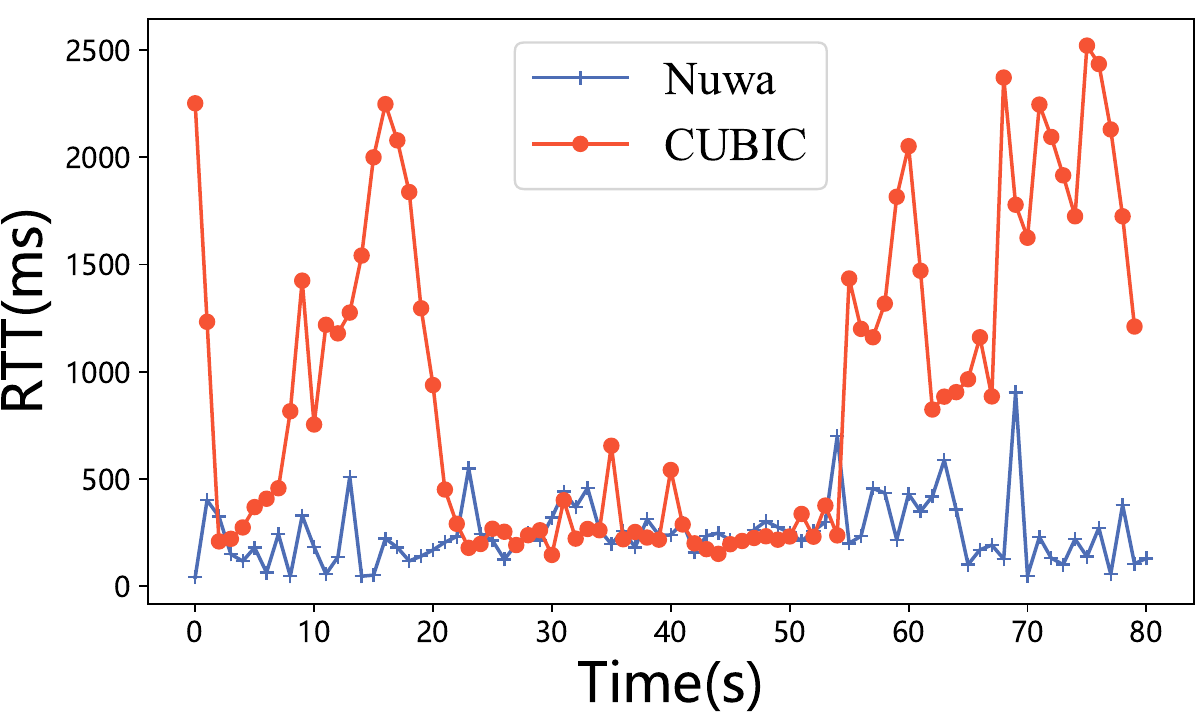}
}%
\subfigure[Queue Delay]{
\includegraphics[width=4.35cm]{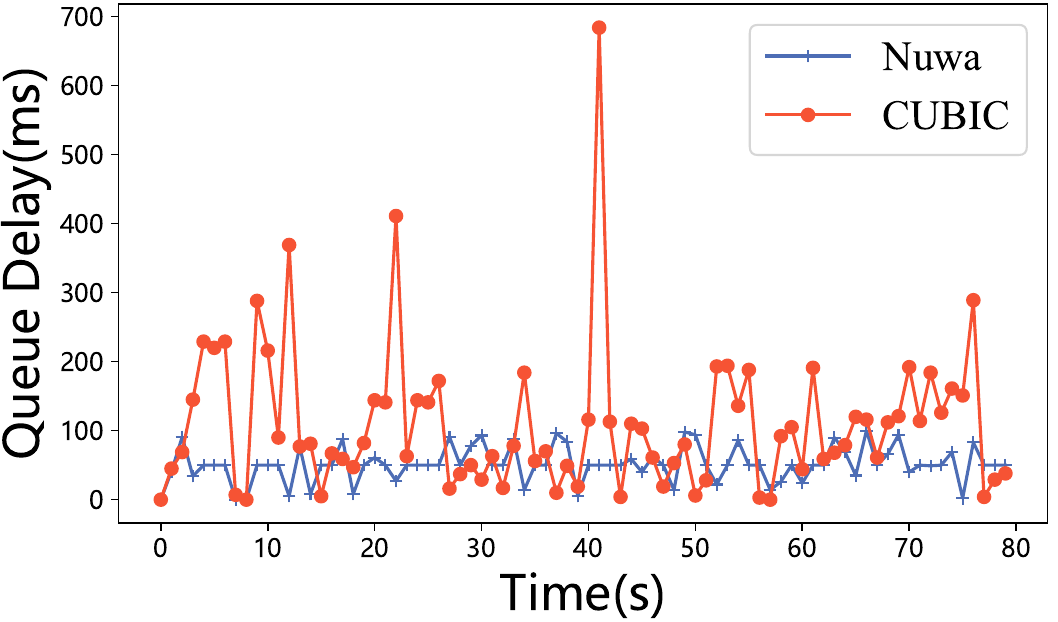}
}%
\subfigure[Throughput]{
\includegraphics[width=4.35cm]{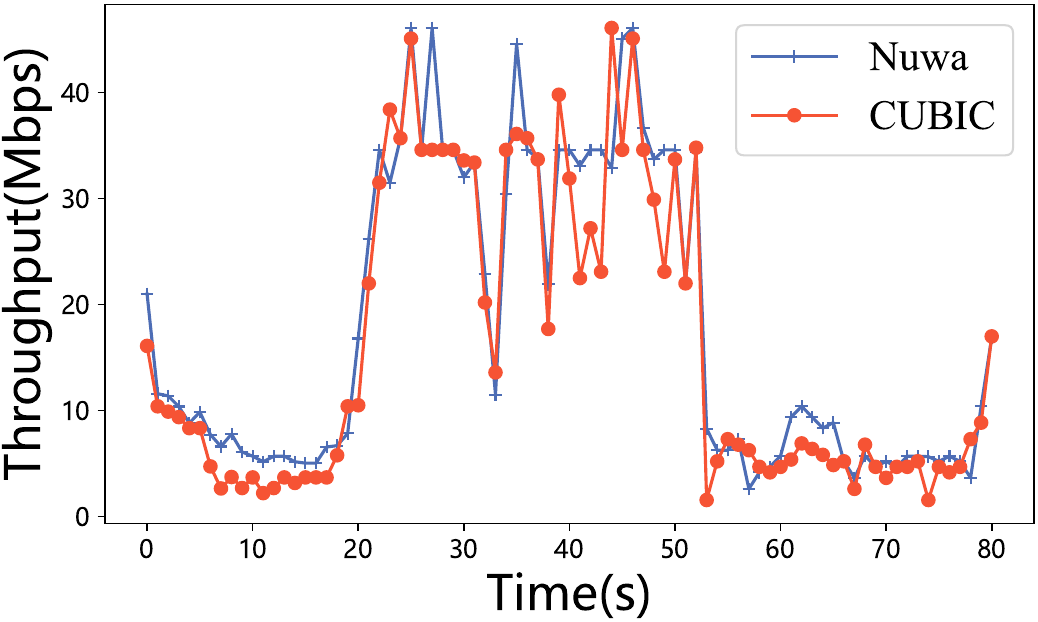}
}%
\caption{Performance comparison of Nuwa algorithm and CUBIC algorithm in 5G network environment.}
\label{fig7}
\end{figure*}

The experimental results are shown in Fig.~\ref{fig7}. The CUBIC algorithm is almost unable to maintain its performance in this network environment. Between 5 and 18 seconds, CUBIC incorrectly estimate the network bandwidth, continue to increase the congestion window, cause network congestion, and cause a significant increase in RTT and queue delay in the network, ultimately resulting in a decrease in data throughput. In the Nuwa algorithm, low RTT and queue delay are almost consistently maintained, which is in stark contrast to the CUBIC algorithm. In addition, we can also see that the window control in Nuwa is closer to the actual changes in the network link, thereby achieving higher throughput. Therefore, we can conclude that Nuwa has good robustness in a fluctuating network environment. It can accurately estimate network changes based on the actual network conditions and make timely adjustments to ensure good performance of data transmission.

\subsubsection{Fairness}
As is well known, delay-based algorithms are prone to significant throughput reduction when competing with loss-based algorithms. In order to avoid this situation, Nuwa sets different target delay $T_{d}$ for each application, and the setting of target delay determines Nuwa's bandwidth usage ability. Therefore, setting a reasonable target delay helps Nuwa obtain fair bandwidth. As shown in Eq.~\ref{eq2}, the sign of $x$ determines whether the window should be increased or decreased, the size of $x$ affects the magnitude of the window adjustment, and an increase in $T_{d}$ can improve Nuwa's competitiveness in the network.

We use the topology shown in Fig.~\ref{fig5} to evaluate the fairness of Nuwa. Since the CUBIC algorithm is the default algorithm in Linux and widely used, we mainly test whether Nuwa and CUBIC could achieve bandwidth fairness. We choose to run stable traces on Cellsim with a total bandwidth of about 10Mbps. We use the iperf tool to analyze the bandwidth usage of the algorithms. For the current network scenario, we set the $T_{d}$ of the Nuwa algorithm to 512. We add a flow to the network every 10 seconds: two Nuwa flows and two CUBIC flows in total.

\begin{figure}
    \centering
    \includegraphics[width=1\linewidth]{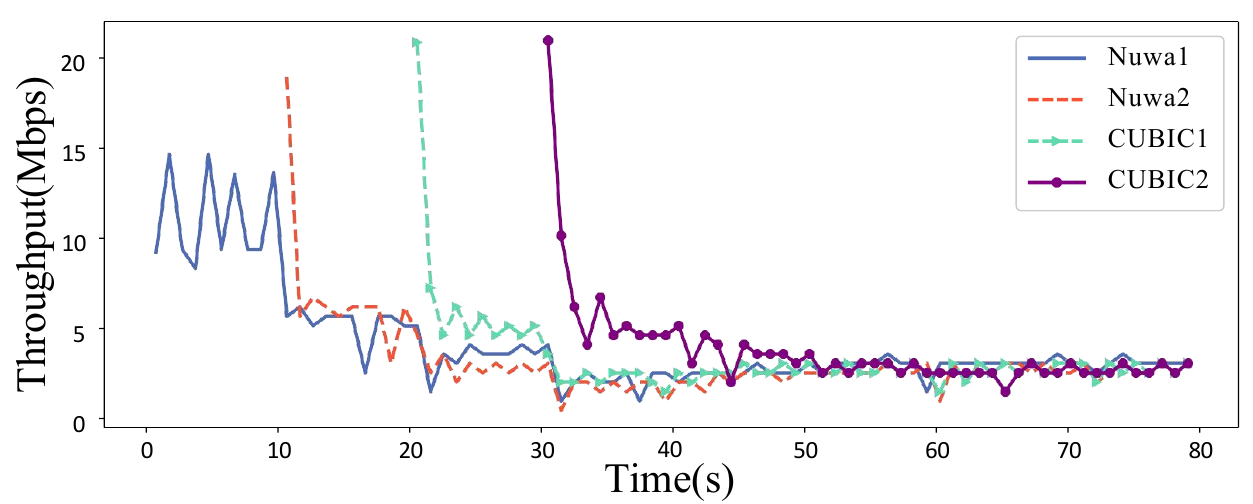}
    \caption{Fairness test of Nuwa algorithm.
}
    \label{fig14}
\end{figure}

The experimental results are shown in Fig.~\ref{fig14}. First, it can be seen that Nuwa achieves fairness within the protocol. When a new Nuwa flow is added, a single Nuwa flow can quickly occupy bandwidth, and two Nuwa flows can share bandwidth. Secondly, Nuwa and CUBIC can share bandwidth. When a CUBIC flow is added, it initially occupies more bandwidth, but at around 50 seconds, Nuwa and CUBIC converge to the same level, thereby sharing the bandwidth resources in the link.

\subsubsection{Preliminary Results}

We set $k=7$ and $T_{d}=750$ and test the transmission performance of Nuwa and other algorithms in the 5G Trace1 network environment. The baseline algorithms include BBR, BBRPlus, CUBIC, Inigo\cite{a33:L}, PCC, Vegas, and Westwood\cite{a35:L}. The experiment is conducted for 80 seconds and the throughput, packet loss rate, and average queue length are used as evaluation metrics. Table.~\ref{tab} shows the average results of multiple experiments. It can be seen that Nuwa can maintain high total data transmission volume and low delay in fluctuating wireless network environments. Compared with other algorithms, Nuwa can increase total data transmission volume by about 4\%$\sim$15.4\% and reduce average queue delay by about 6.9\%$\sim$29.4\%. At the same time, Nuwa algorithm can also maintain a lower packet loss rate.

\begin{table}
\caption{Performance comparison of different algorithms in 5G}
\setlength{\tabcolsep}{0.5mm}{
\begin{tabular}{|c|c|c|c|}
\hline
\textbf{Algorithm} & \textbf{Transfered data size} & \textbf{Packet loss rate} & \textbf{Average queue delay} \\ \hline
BBR                & 168.5MB              & 0.54\%                    & 144.93ms                      \\
BBRPlus            & 172.3MB              & 0.61\%                    & 137.48ms                      \\
CUBIC              & 187.1MB              & 0.87\%                    & 165.79ms                      \\
Inigo              & 177.5MB              & 0.71\%                    & 146.3ms                       \\
PCC                & 185.3MB              & 1.02\%                    & 177.5ms                       \\
Vegas             & 179.1MB              & 0.82\%                    & 169.47ms                      \\
Westwood           & 177.8MB              & \textbf{0.51\%}           & 134.74ms                      \\
Nuwa               & \textbf{194.5MB}     & 0.56\%                    & \textbf{125.4ms}              \\ \hline
\end{tabular}}
\label{tab}
\end{table}

\section{Improving Nuwa with Reinforcement Learning}

Nuwa, as a heuristic congestion control algorithm, can accurately judge the network bandwidth changes by observing real network feedback in specific network environments, maintain good performance in case of network fluctuations, and effectively prevent throughput degradation caused by sudden network degradation. However, when the network environment changes, old parameters may no longer be suitable for the current network environment, which may affect the transmission performance of Nuwa.

\begin{figure}[!t]
\centering
\subfigure[$k$=9]{
\includegraphics[width=8.7cm]{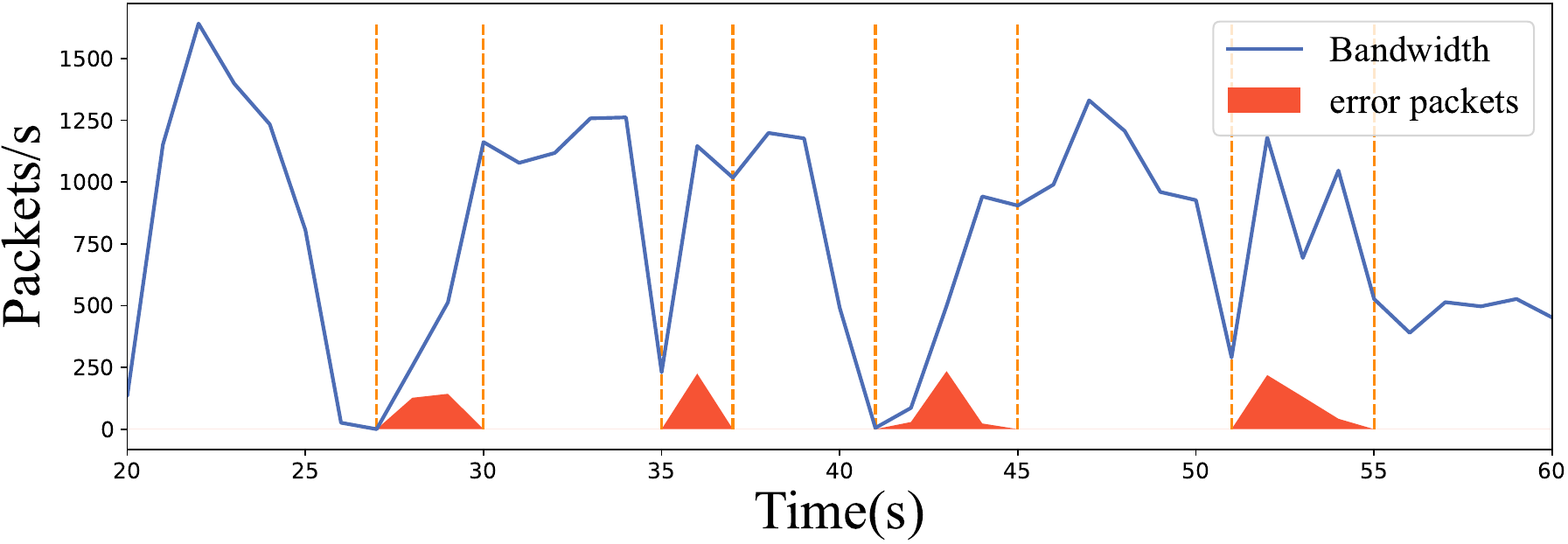}
}%

\subfigure[$k$=1]{
\includegraphics[width=8.7cm]{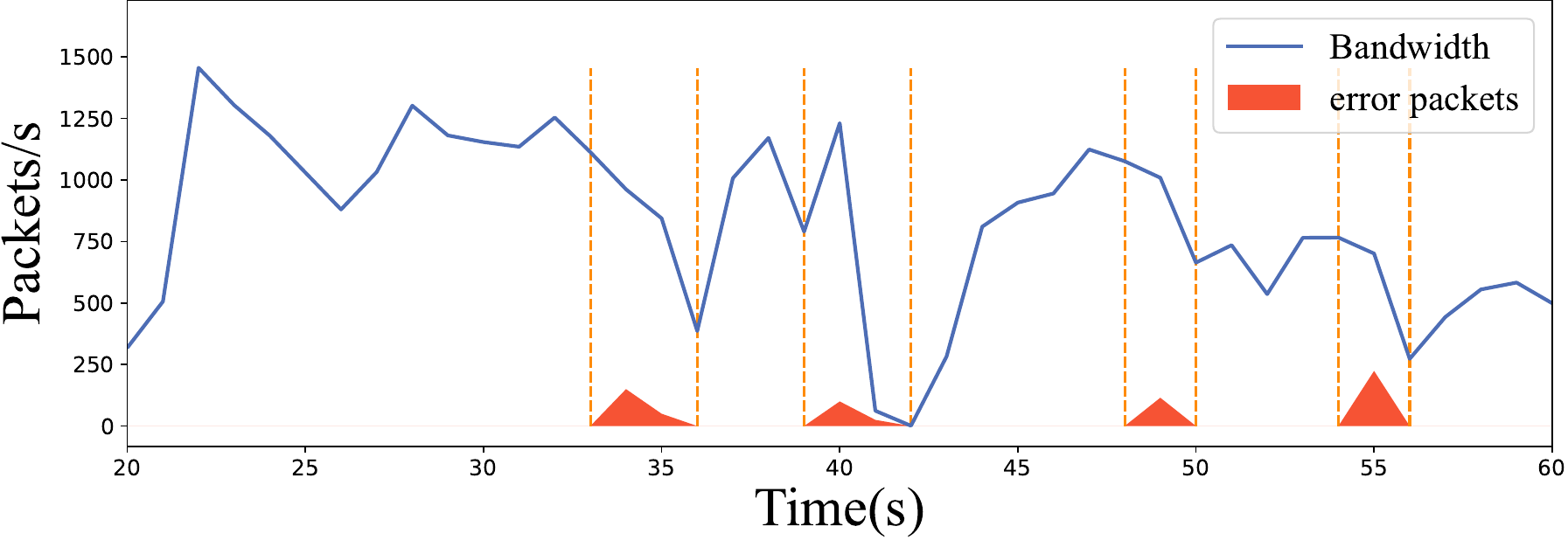}
}%
\caption{Effect of different $k$ values on the performance of Nuwa.}
\label{fig8}
\end{figure}

The parameter $k$ is a crucial factor controlling the window adjustment magnitude in Nuwa. The window size adjustment must meet the needs of the network status. When the window size is too large, it may result in packet loss when the number of sent packets exceeds the capacity of the link. When the window size is too small, it may lead to a decrease in throughput and insufficient utilization of bandwidth resources. Therefore, the selection of $k$ value directly affects the efficiency of data transmission. The range of $k$ value in the algorithm is [1-9]. We conduct experiments with different $k$ values, and the results are shown in Fig.~\ref{fig8}. It can be observed that the larger the $k$ value, the more sensitive the algorithm is to network fluctuations and the more affected the throughput is. When the network fluctuates, the throughput will significantly decrease. Meanwhile, smaller $k$ values will result in the window changes lagging behind the network changes, leading to packet loss at the falling edge. 

In the preliminary performance testing results of Nuwa, we believe that $k=7$ is suitable for the 5G network environment. Compared to 4G, 5G uses higher frequency radio spectrum, further enhancing data transmission capabilities. However, high-frequency signal transmission distances are shorter, and to achieve better coverage, the 5G network require more base stations and devices, increasing network construction and operation costs. Therefore, 4G and 5G coexist in current networks, that is the network we use usually switch back and forth between 4G and 5G in everyday life. $ k=7$ may not be suitable for the 4G network environment. We must consider the differences between different network environments and make different decisions for different network environments. To validate our idea, we conduct throughput experiments in real 4G and 5G networks using the topology structure shown in Fig.~\ref{fig1}, to test the impact of different $k$ values on the performance of the Nuwa algorithm in different network environments. Meanwhile, we use CUBIC as a benchmark algorithm to verify the effectiveness of Nuwa.

\begin{figure}
    \centering
    \includegraphics[width=1\linewidth]{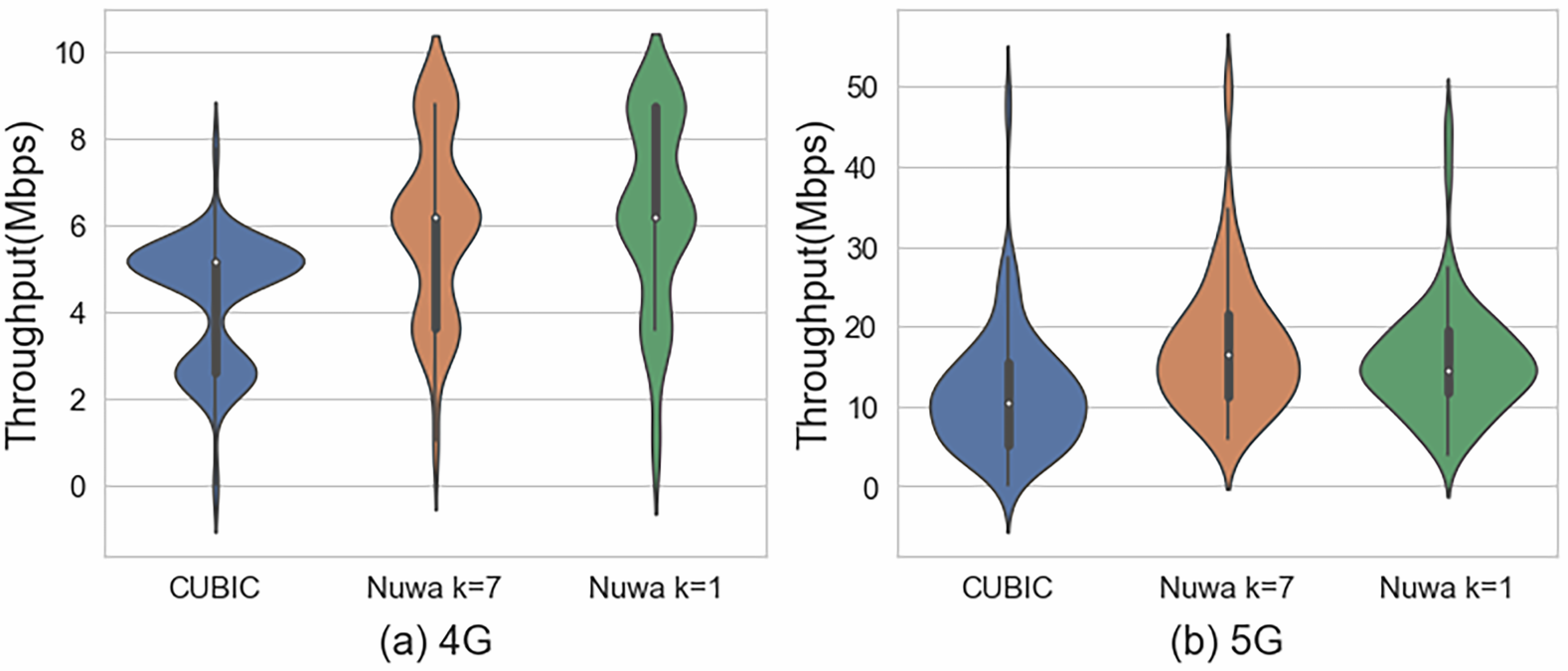}
    \caption{Throughput performance of Nuwa algorithm with different settings and CUBIC in 4G and 5G networks.
}
    \label{fig9}
\end{figure}

Experimental results are shown in Fig.~\ref{fig9}. Firstly, we can observe that, compared to the 4G network, the three algorithms have higher and more concentrated throughput in the 5G network, indicating that there are indeed differences between 4G and 5G networks, and different network characteristics lead to different data transmission capabilities of the same congestion control algorithm. Secondly, regardless of whether it is in 4G or 5G networks, Nuwa can effectively improve the throughput of the original algorithm. In addition, in the 4G network, when $k=1$, Nuwa has a higher overall throughput level and better transmission performance. While in the 5G network, $k=7$ leads to higher throughput for Nuwa. Therefore, we conclude that Nuwa can indeed improve the transmission performance of the original algorithm, but for different network environments, we must re-select suitable parameters according to the network characteristics to enhance Nuwa's adaptability to the network environment. $k$ is a key parameter that controls the aggressiveness of the window change in the Nuwa algorithm. We believe that using reinforcement learning to adjust the $k$ parameter can make the Nuwa algorithm more adaptable to a wider range of network environments.

\begin{figure}
    \centering
    \includegraphics[width=1\linewidth]{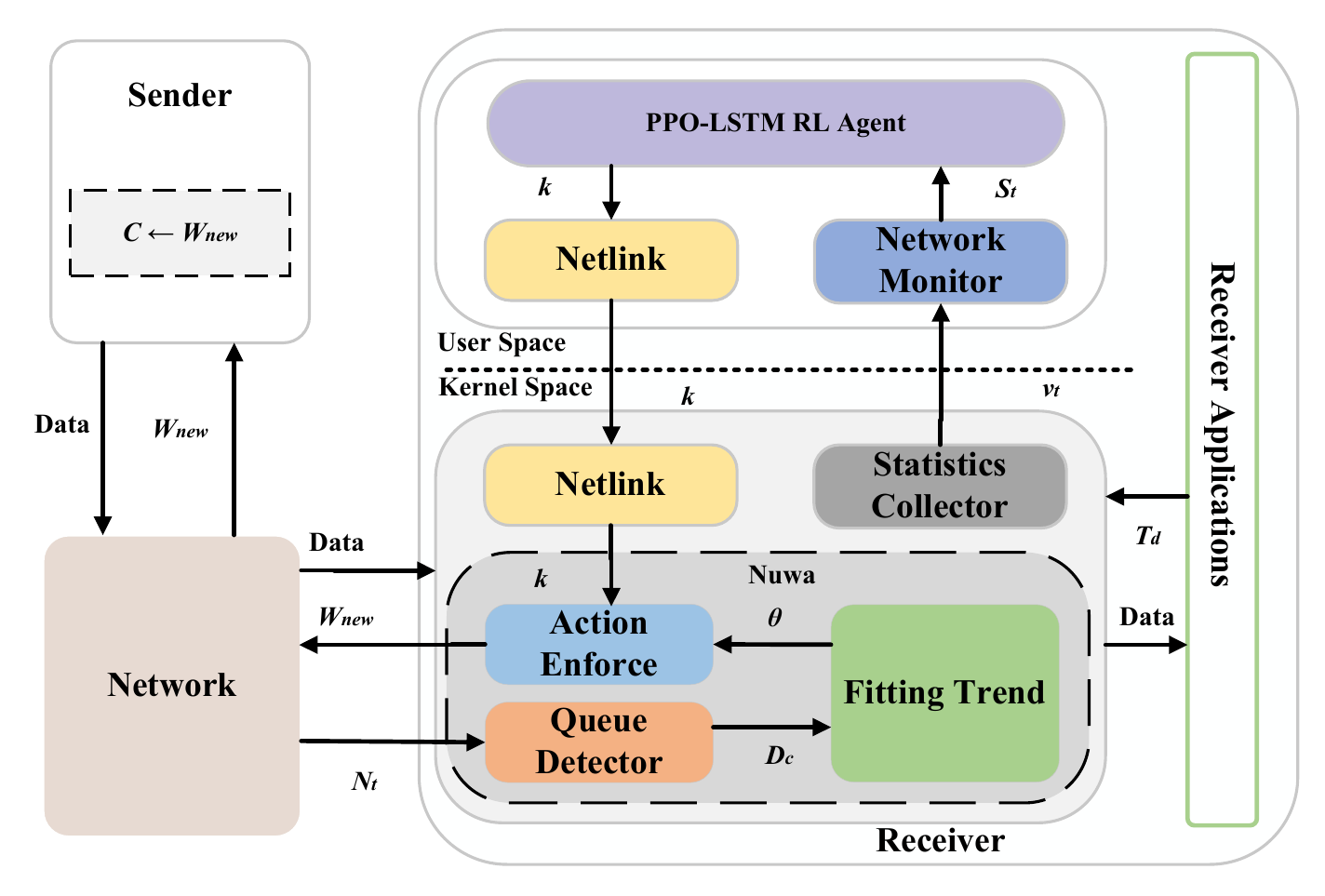}
    \caption{Improving Nuwa with reinforcement learning.
}
    \label{fig10}
\end{figure}

We combine reinforcement learning with Nuwa, and the overall architecture is shown in Fig.~\ref{fig10}. We deploy the Nuwa algorithm in the kernel of the receiver-side, and we deploy the reinforcement learning component in the application layer of the receiver-side. As the kernel state and user state cannot communicate directly in Linux, we redesign the data collection and transmission parts. We use eBPF technology to implement the Statistics Collector module for real-time collection of kernel data. In short, eBPF collects kernel data through kprobe technology and forms a vector $v_{t}$ representing the current network state, which is then stored in the eBPF Map. User programs can retrieve the kernel data $v_{t}$ through system calls. The Network Monitor in the user layer receives $v_{t}$ and forms a network state vector $S_{t}$ containing historical states, which is passed on to the RL Agent. We use Netlink to transfer user layer data to the kernel, mainly referring to the transmission of the action $a_{t}$ generated by the RL Agent, i.e., the $k$ value most suitable for the current network state selected by the RL Agent for the Nuwa algorithm in the kernel. The Nuwa algorithm calculates the window value $w_{new}$ based on the specified $k$ value and the target delay $T_{d}$ set by the application, and sends it through ACK to the sender-side. 

\subsection{Why PPO-LSTM}

\begin{figure*}[!t]
\centering
\subfigure[A2C]{
\includegraphics[width=8.7cm]{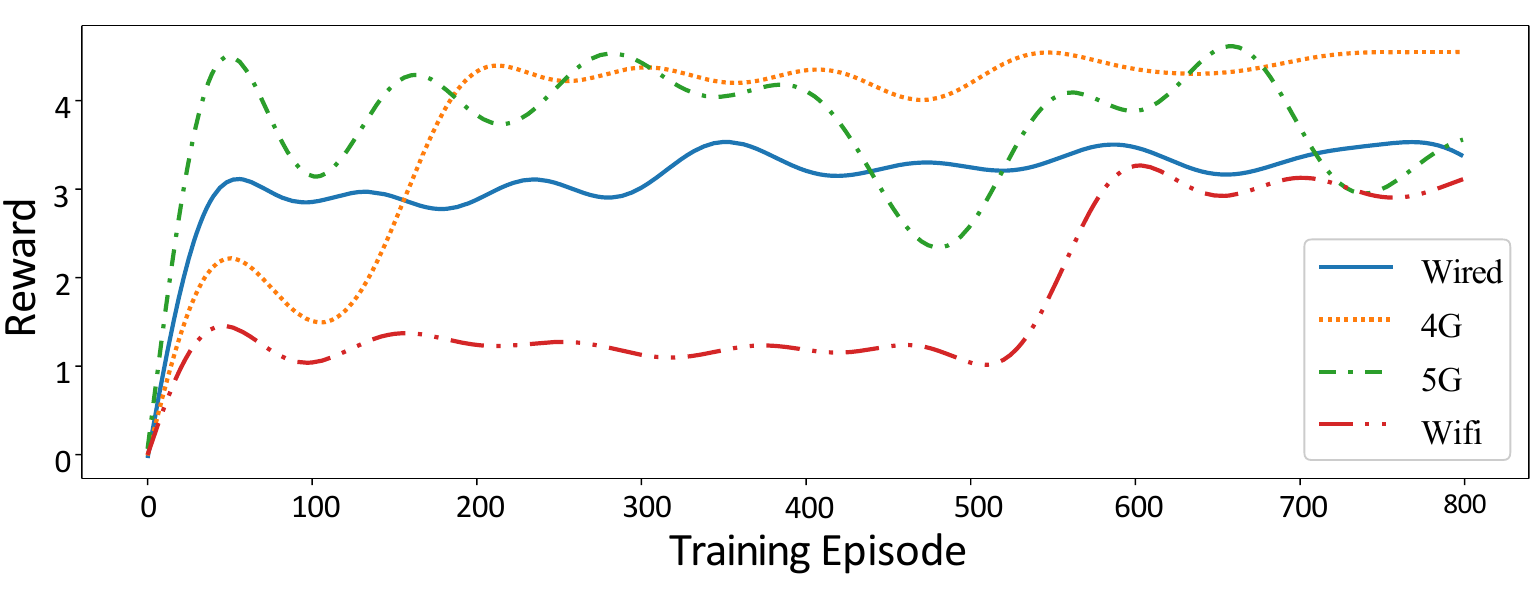}
}%
\subfigure[DQN]{
\includegraphics[width=8.7cm]{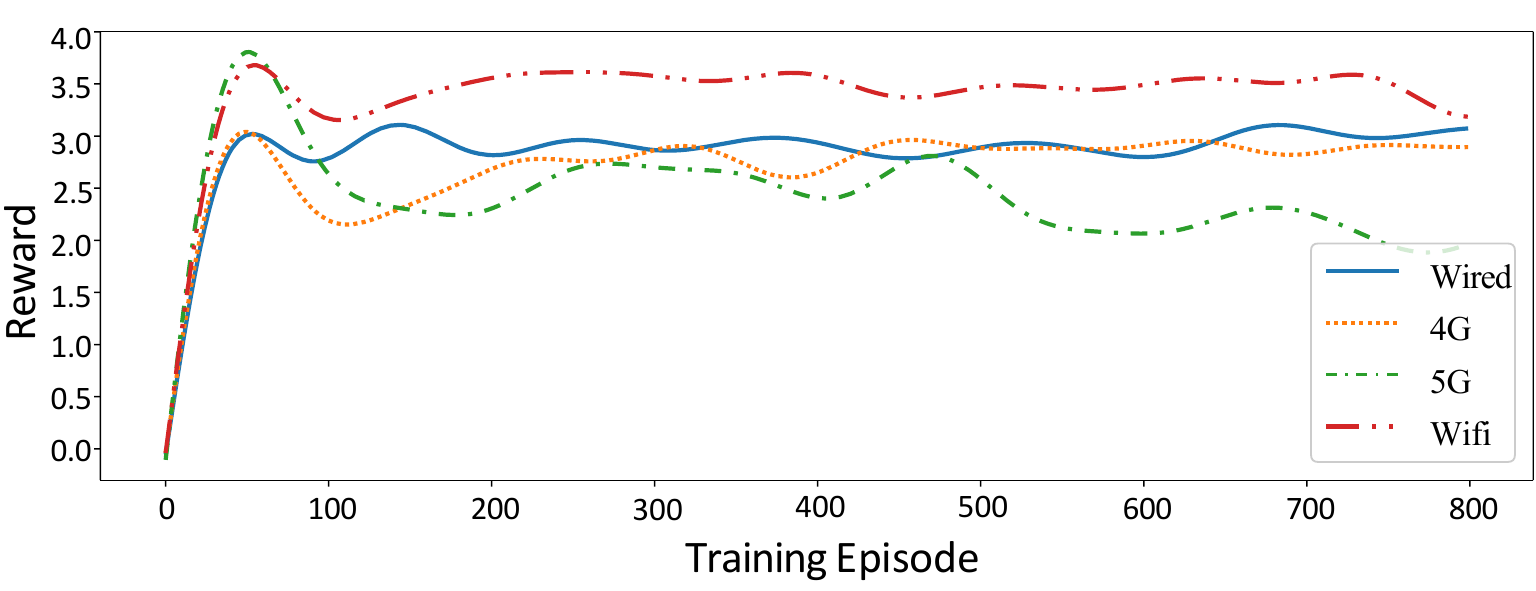}
}%
\vspace{-2mm}

\subfigure[PPO]{
\includegraphics[width=8.7cm]{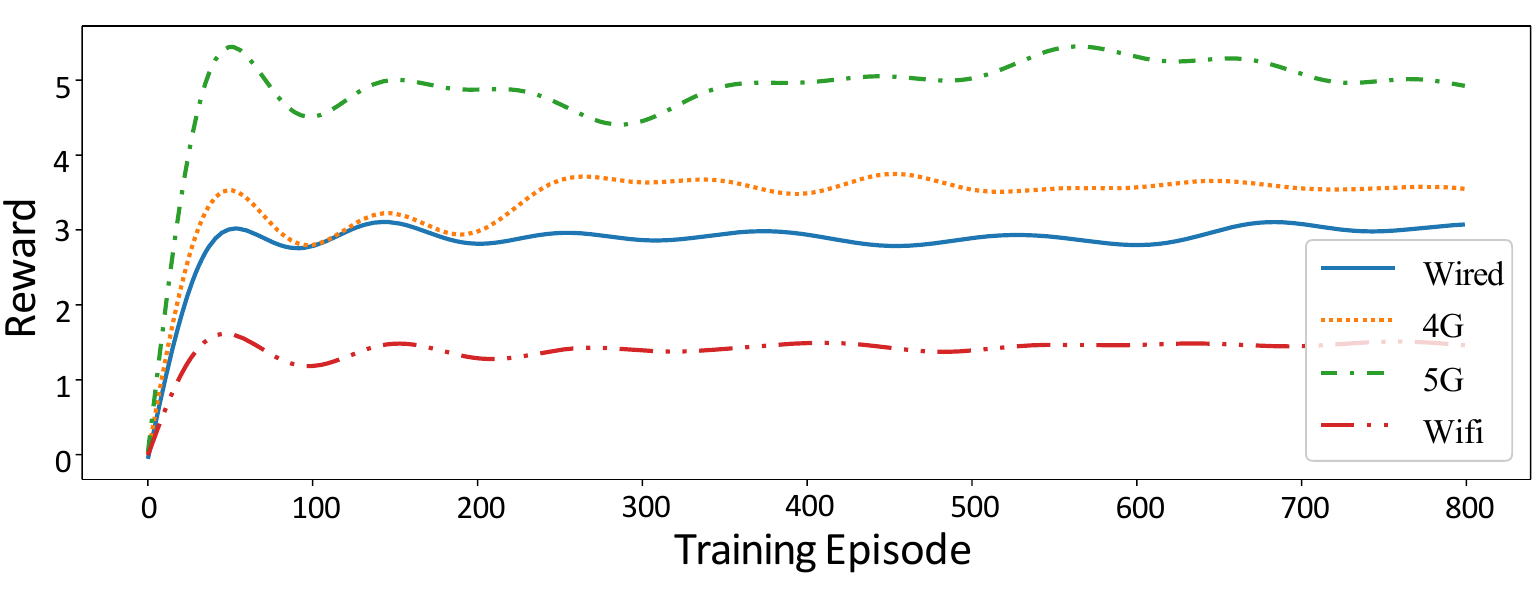}
}%
\subfigure[PPO-LSTM]{
\includegraphics[width=8.7cm]{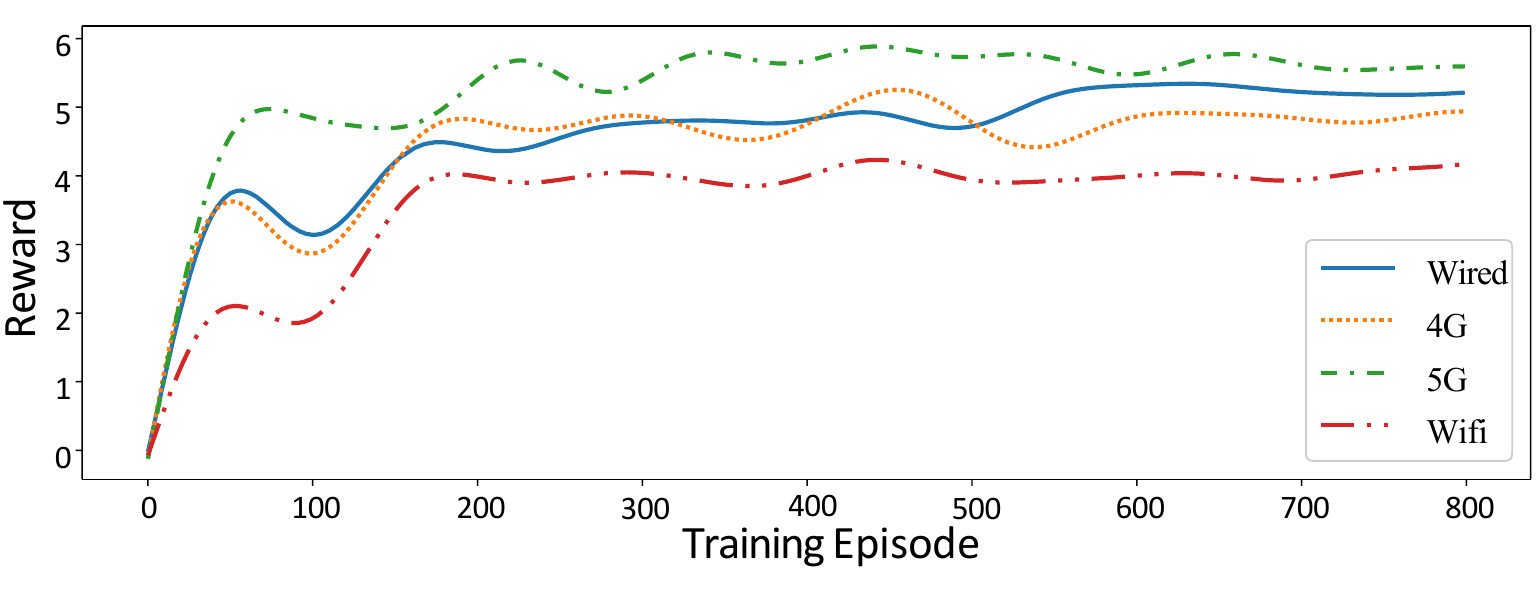}
}%
\caption{Performance of each reinforcement learning algorithm in different network environments.}
\label{fig11}
\end{figure*}

We select four state-of-the-art reinforcement learning algorithms, A2C\cite{a12:V}, DQN\cite{a13:V}, PPO\cite{a14:J}, and PPO-LSTM\cite{a15:S}, and compare their performance in wireless networks. We aim to achieve a good balance between higher reward values and faster learning efficiency. Using the experimental topology described in section~\ref{sec}, we conduct experiments in wired networks, Wifi, 4G, and 5G, and the results are shown in Fig.~\ref{fig11}.

We first observe that the same reinforcement learning algorithm exhibits performance differences in different network environments, especially for A2C and PPO. Secondly, in the wired network, the reward values of the four algorithms can almost stabilize at a moderate level in a short time, with relatively small differences between each other. However, in wireless networks, the performance differences between each algorithm are much more significant. In Wifi, the learning efficiency of A2C is very low, and it takes about 500 rounds of training to learn to a stable level. Although the learning efficiency of PPO algorithm is outstanding, the overall reward value is very low. In contrast, the performance of DQN and PPO-LSTM is much better, especially DQN balances well between learning efficiency and reward value. In mobile networks, under the condition that the learning efficiency is not much different, PPO and PPO-LSTM can obtain higher levels of reward values than A2C and DQN, and PPO-LSTM has smaller performance differences in 4G and 5G networks, indicating that PPO-LSTM is more stable. Therefore, we choose PPO-LSTM as the reinforcement learning algorithm to improve Nuwa.

PPO-LSTM is an improvement on PPO. PPO involves learning and computing updates using a minimum clipping strategy during each iteration, and the significant difference between the new and old policies can easily lead to incorrect decisions. PPO-LSTM utilizes LSTM\cite{a37:S} to construct a policy network. Thanks to LSTM's powerful ability to handle time-series data, combining LSTM with PPO allows the model to have memory, effectively improving the model's performance.

The training process of reinforcement learning algorithm can be simply described as follows: the policy network generates a policy $p_{t}$, and the agent selects the next action $a_{t}$ based on the observed state $s_{t}$ and the existing policy $p_{t}$, in the hope of obtaining the maximum expected reward $r_{t}$. PPO-LSTM is still based on the Actor-Critic (AC) network architecture. It consists of two Actor networks, one responsible for interacting with the environment to generate new policies, and one that stores the old policies and waits to be optimized. It also includes a Critic network, responsible for evaluating policies. By continuously updating the policy, reinforcement learning will eventually find the optimal policy $\pi$ with parameters $\theta$ to maximize the total reward $J_{PPO-LSTM}(\theta )=E[min(r(\theta )\hat{A}_{t},clip(r(\theta),1-\epsilon,1+\epsilon)\hat{A}_{t}  ) ]$, where $r(\theta )=\frac{\pi _{\theta}\left ( a_{t} |s_{t}  \right )  }{\pi _{\theta _{old} } }$, $\epsilon$ is a hyperparameter.

The Critic network uses an advantage function to evaluate the performance of the selected action. The advantage function can be expressed as:
\begin{equation}
    \label{eq12}
	\hat{A}_{t}=\delta _{t}+(\lambda \gamma )\delta _{t+1}+\dots +(\lambda \gamma )^{T-t+1} \delta _{T-1}
\end{equation}
where
\begin{equation}
	\label{delta-t}
	\delta _{t}=r_{t}+\gamma V_{\phi }(s_{t+1})-V_{\phi }(s_{t})
\end{equation}
$\gamma$ is a trade-off factor between bias and variance in the generalized advantage estimation. And $V_{\phi }(s)$ is state–value function defined as:

\begin{equation}
	\label{delta-t}
	V_{\phi }(s) = E_{\pi}[\sum_{k=1}^{\infty } \gamma ^{k}r_{t+k+1}|s_t=s ]
\end{equation}
Algorithm~\ref{alg3} describes the PPO-LSTM algorithm.

\begin{algorithm} 
	\caption{Reinforcement Learning Algorithm} 
	\label{alg3} 
	\begin{algorithmic}[1]
		\STATE{Input: Initialize policy $\pi$ with parameters $\theta_0$, initialize value function parameters $\phi_0$;}
		\FOR{$k\in \left \{1,2,\ldots\right \}$} 
		\STATE Run policy $\pi_{\theta}$ for T timesteps:
		\STATE \qquad Select action $a_t$;
		\STATE \qquad Execute $p_{t}$ to the steady phase according to Equation (\ref{eq7});
		\STATE \qquad Observe the new states $s_{t+1}$ and the reward $r_t$;
		\STATE \qquad Store sample  $\tau_i = \left \{s_t, a_t, r_t\right \}$ into replay buffer;
		\STATE Collect set of trajectories  $D_k = \left \{\tau_i\right \}$;
		\STATE Compute advantages $\hat{A_t}$;
		\STATE $r(\theta) =\frac{\pi _{\theta}\left ( a_{t} |s_{t}  \right )  }{\pi _{\theta _{old} } }$;
		\STATE $\pi _{\theta _{old} } \gets \pi_{\theta}$;
		\STATE Update the policy by maximizing the objective function $J_{PPO-LSTM}(\theta)$;
		\STATE Update policy parameters $\theta$;
		\STATE Update value function parameters $\phi$;
		\ENDFOR 
	\end{algorithmic} 
\end{algorithm}

\subsection{The RL Agent}

\textbf{State}. The state information in PPO-LSTM is used for action inference and model learning. The Network Monitor generates the state vector $v_{t}$ based on the statistical data collected from the kernel. $v_{t}$ must contain sufficient network state information for the RL agent to make correct decisions based on this information. We select the following parameters for the state space:

$g_{t}$:The total amount of data received by receiver-side within a certain period of time.

$w_{t}$:The estimated window size of the receiver-side.

$r_{t}$:The size of the real-time buffer cached by receiver-side.

$l_{t}$:The one-way delay of the last transmitted data packet.

RL agent needs to not only focus on the current state, but also learn from previous historical states. Using historical knowledge can enable the RL agent to make decisions that better align with the trend of network states, effectively improving the performance of the algorithm. Therefore, the state at time $t$ is defined as Eq.~\ref{eq4}, where $n$ is the historical length taken into consideration.

\begin{equation}
S_{t}=\left ( v_{t-n}, v_{t-n+1},...,v_{t}\right )  \label{eq4} 
\end{equation}

\textbf{Action}. Action $a_{t}$ mainly refers to adjusting the parameter $k$ in the Nuwa algorithm. We use the product of the action and $k$ to adjust $k$ to achieve dynamic adaptation of the sending window to the network. The BBR\cite{a16:N} algorithm suggests using a behavior approximating 2.89 as one of the actions to allow the agent to quickly adapt to changes in the network environment. Inspired by BBR, to achieve the optimal sending rate, we make small adjustments to the $k$ value using 1.25 and 1.05, respectively. After the algorithm reaches stability, $action=1$ can maintain stable behavior at the optimal operating point. When the network condition begins to deteriorate, we use $action=0.25$ to achieve reverse adjustment of the window. In summary, based on the range of the window, we divide the actions output by the RL into five types, and the action space of the RL agent is shown below:

\begin{equation}
Action=\left \{ 0.25, 1, 1.05, 1.25, 2.85 \right \}  \label{eq5} 
\end{equation}

\textbf{Reward}. Alpha fairness function\cite{a17:R} is widely used in evaluating the quality of bandwidth allocation when multiple flows share the same link in resource allocation. The alpha fairness function is defined as follows:

\begin{equation}
    \label{eq6} 
    U_{\alpha }\left ( x \right ) =\left\{ 
\begin{aligned}
 \log_{}{(x)}, \quad \alpha =1\\
 \frac{x^{1-\alpha } }{1-\alpha }, \quad \alpha > 0, \alpha \ne 1
\end{aligned}
\right.
\end{equation}

RLCC\cite{a18:Z} points out that congestion control algorithms based on reinforcement learning will suffer significant performance degradation when only a single metric is used in the reward function. Meanwhile, considering that Nuwa-RL aims to achieve a trade-off between throughput and latency at receiver-side, while maintaining low latency and low packet loss in the network, as well as achieving high bandwidth utilization and good fairness. Therefore, we consider throughput ($b$), latency ($\tau _{r}$), and packet loss rate ($\tau _{l}$) in the reward function. Combining with the Alpha Fairness function, we set the utility function as shown in Eq.~\ref{eq7}:

\begin{equation}
 U\left ( b(t), \tau_{r}(t), \tau_{l}(t) \right ) =\gamma \cdot U_{\alpha }{( b(t))}-\theta \cdot U_{\alpha }{( \tau_{r}(t))}-\varphi \cdot U_{\alpha }{( \tau_{l}(t))}    \label{eq7} 
\end{equation}

Among them, $\gamma$, $\theta$, and $\varphi$ are constants, representing the importance of throughput, delay, and packet loss rate in the reward function. In the experiment, we set $\gamma=0.4$, $\theta=0.4$, $\varphi=0.2$, and $\alpha=0.6$.

\textbf{Training}. We implement the RL agent using the Stable Baselines3\cite{a50:A} framework. To make the model results closer to the real network environment, we conduct model training and testing on the testbed Cellsim. In Cellsim, we deploy network bandwidth traces collected under real network conditions as the network transmission environment. We set the training period to 800 steps to avoid overfitting.

\subsection{Preliminary Evaluation}
In this section, we conduct preliminary tests on an enhanced version of the Nuwa algorithm that incorporates reinforcement learning(hereinafter referred to as Nuwa-RL). A more detailed evaluation is presented in Section~\ref{Sec}.

\subsubsection{Experimental Settings} \label{sec}
\begin{figure}
    \centering
    \includegraphics[width=1\linewidth]{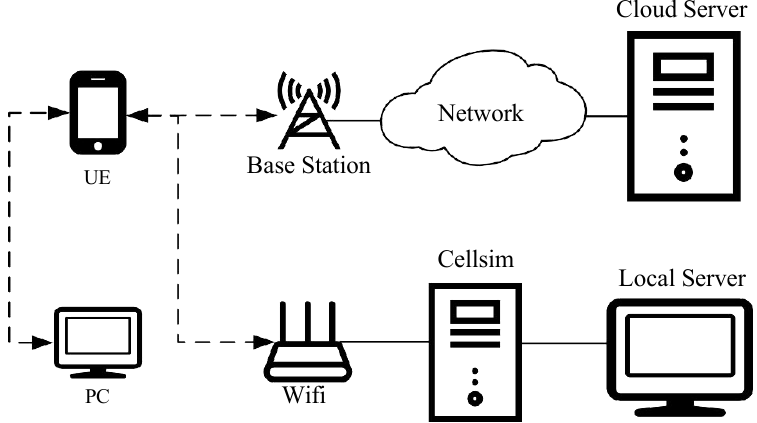}
    \caption{Topology of the experiment of Nuwa-RL.
}
    \label{fig12}
\end{figure}

The experimental environment includes both real-world and simulated environments, as shown in Fig.~\ref{fig12}. The receiving end PC used is a Think Pad X1 Carbon laptop, where the Nuwa-RL algorithm is deployed internally and connected to the wireless network via a mobile phone. For local network emulation experiments, we use a large machine (DELL i5 4G) and deploy Cellsim network emulation internally. At the same time, we use another large machine as a local server and deploy cloud services using services provided by Huawei Cloud (4vCPU, 8GiB). We deploy the same data on the cloud server and local server to ensure the consistency of experimental results.

\subsubsection{Preliminary Results}
\begin{figure}[!t]
\centering
\subfigure[Wired]{
\includegraphics[width=4.3cm]{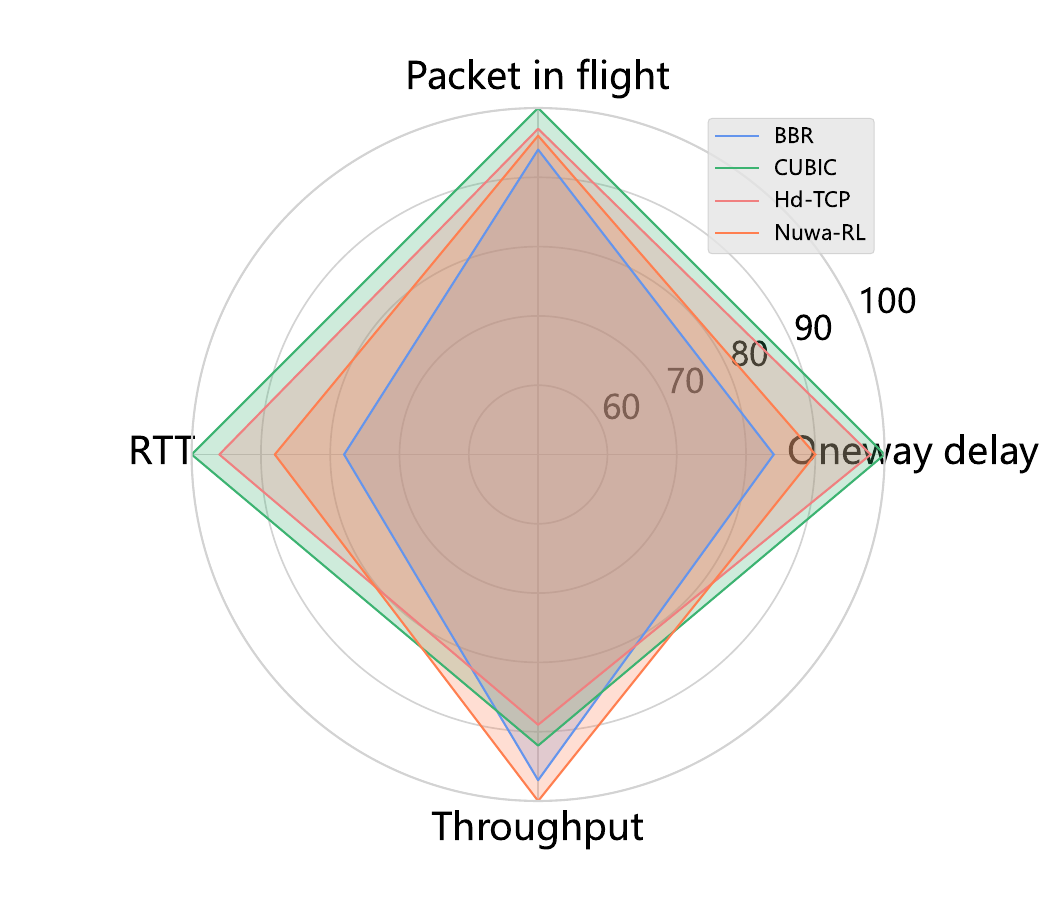}
}%
\subfigure[Wifi]{
\includegraphics[width=4.3cm]{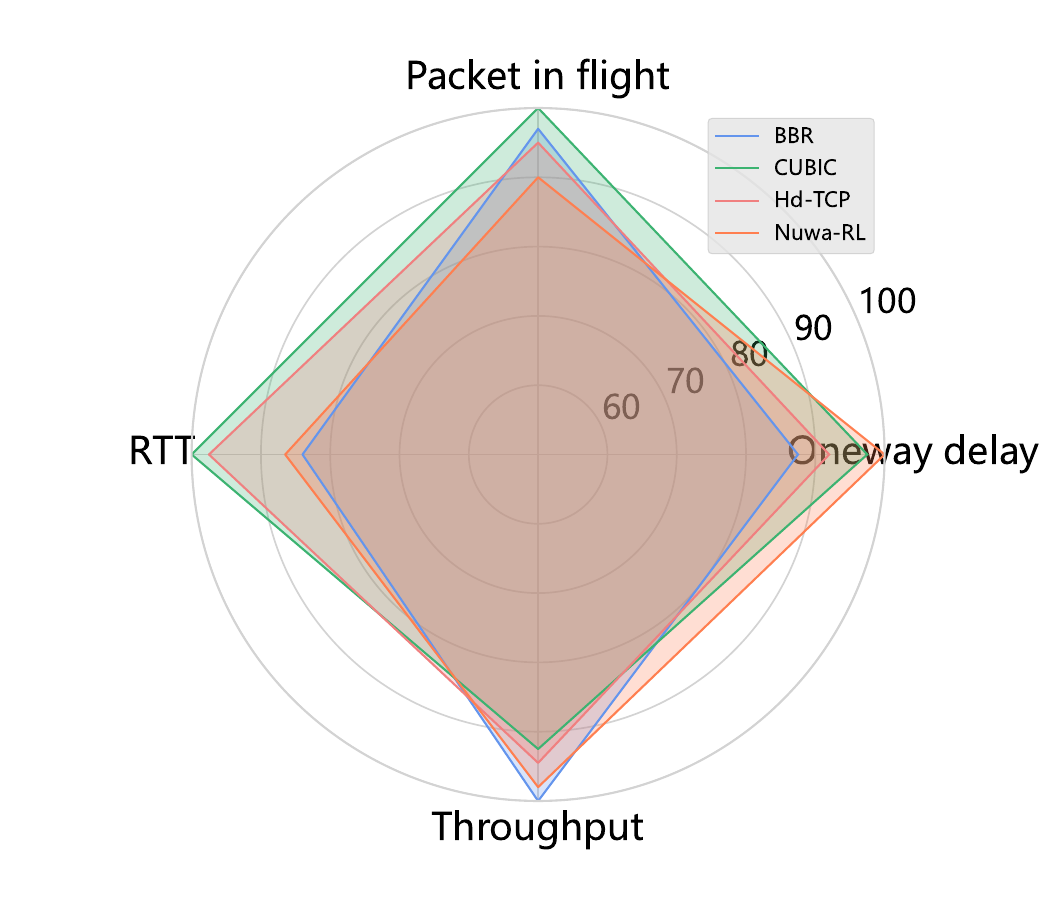}
}%

\subfigure[4G]{
\includegraphics[width=4.3cm]{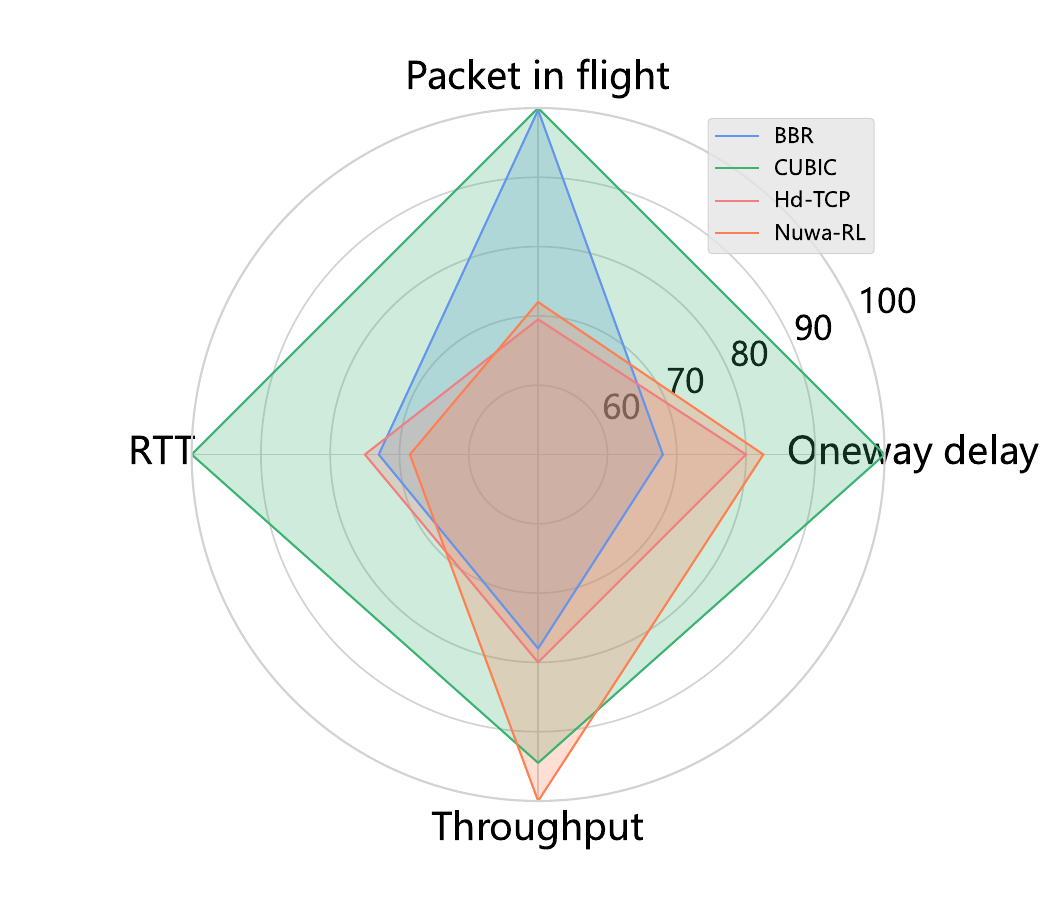}
}%
\subfigure[5G]{
\includegraphics[width=4.3cm]{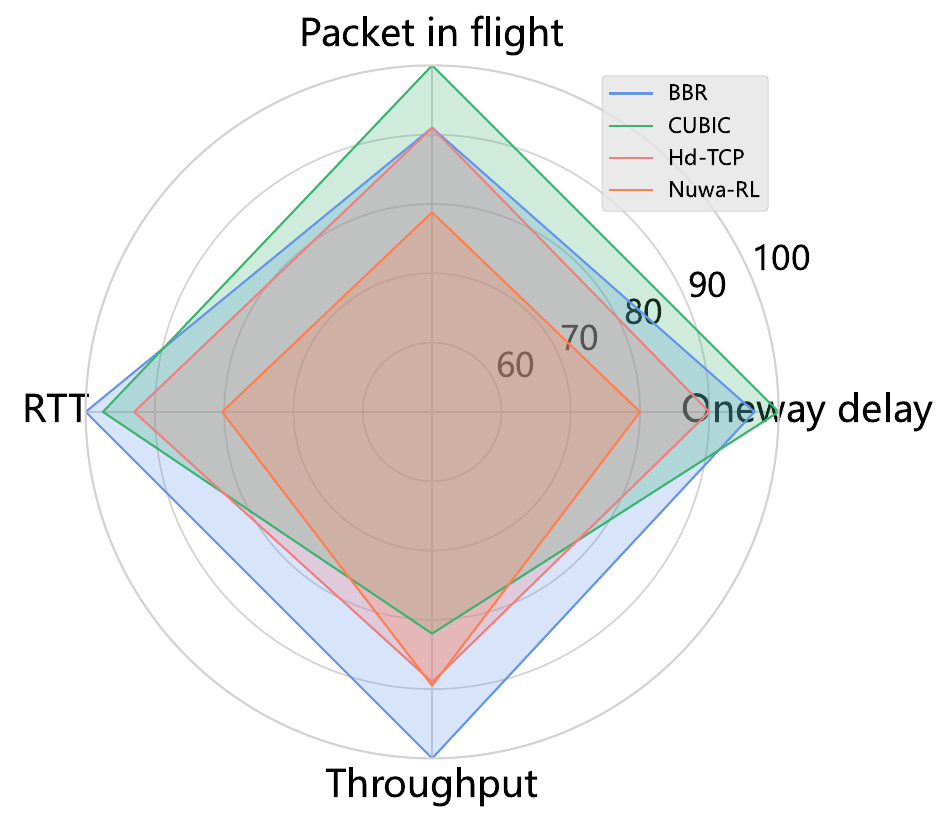}
}%
\caption{Preliminary performance test results of Nuwa-RL algorithm.}
\label{fig13}
\end{figure}

We compare BBR, CUBIC, Hd-TCP\cite{a19:L}, and Nuwa-RL in our test, with Hd-TCP being a reinforcement learning algorithm. We test them in real network environments, including wired, Wifi, 4G, and 5G. The experiment is conducted in the form of file transfer, with each experiment lasting for 40 seconds. We choose packet flight, one-way delay, RTT, and average throughput as the indicators for algorithm evaluation. Each algorithm is tested multiple times in different networks, and the results are averaged. In addition, to better display the performance differences between the algorithms, we divide each indicator by the maximum value of the corresponding indicator in the four algorithms in the same network environment. The experimental results are shown in Fig.~\ref{fig13}.

From the experimental results, we find that BBR, as a delay-based algorithm, has advantages in relatively stable network transmission environments. It can maintain low transmission delay and queue delay in the network while having high data transmission capacity. However, in wireless networks, BBR obviously struggles to maintain its performance, with very low throughput in 4G network environments and significant delay in 5G network environments. CUBIC uses packet loss as a congestion signal, and its inherent characteristics cause it to send too many packets to the link at the beginning, resulting in high data transmission delay despite having high throughput. Hd-TCP can achieve performance similar to CUBIC in a stable network environment, but in 4G and 5G network environments, Hd-TCP has more advantages and can maintain lower delay thanks to the learning of the RL agent. Compared to the other three algorithms, the Nuwa-RL algorithm achieves better data transmission capabilities in multiple network environments. In wired networks, Nuwa-RL achieves throughput similar to CUBIC while providing queue delay control second only to BBR. In 4G networks, Nuwa-RL achieves the highest throughput and lowest RTT, while in 5G networks, it provides the lowest delay while maintaining good throughput. In summary, Nuwa-RL demonstrates good environmental adaptability and can effectively control delay while providing decent throughput capabilities in multiple network environments.

\subsubsection{Fairness}
In Nuwa-RL, the fairness is ensured by selecting an appropriate value of k and using the alpha-fair function. Updating the value of $k$ when the available bandwidth changes can effectively adjust the magnitude of window changes and achieve the goal of quickly modifying the window. Applying the alpha-fair function in the reward function can avoid excessively high reward values in multi-flow environments that lead to bandwidth monopolization. We conduct multi-flow transmission experiments in the local emulation environment shown in Fig.~\ref{fig12}. We still use CUBIC to test the fairness of Nuwa-RL. The experiment add a new flow to the network every 10 seconds for the same bottleneck link: two Nuwa-RL flows and two CUBIC flows.

\begin{figure}
    \centering
    \setlength{\abovecaptionskip}{-0.1cm}
    \includegraphics[width=1\linewidth]{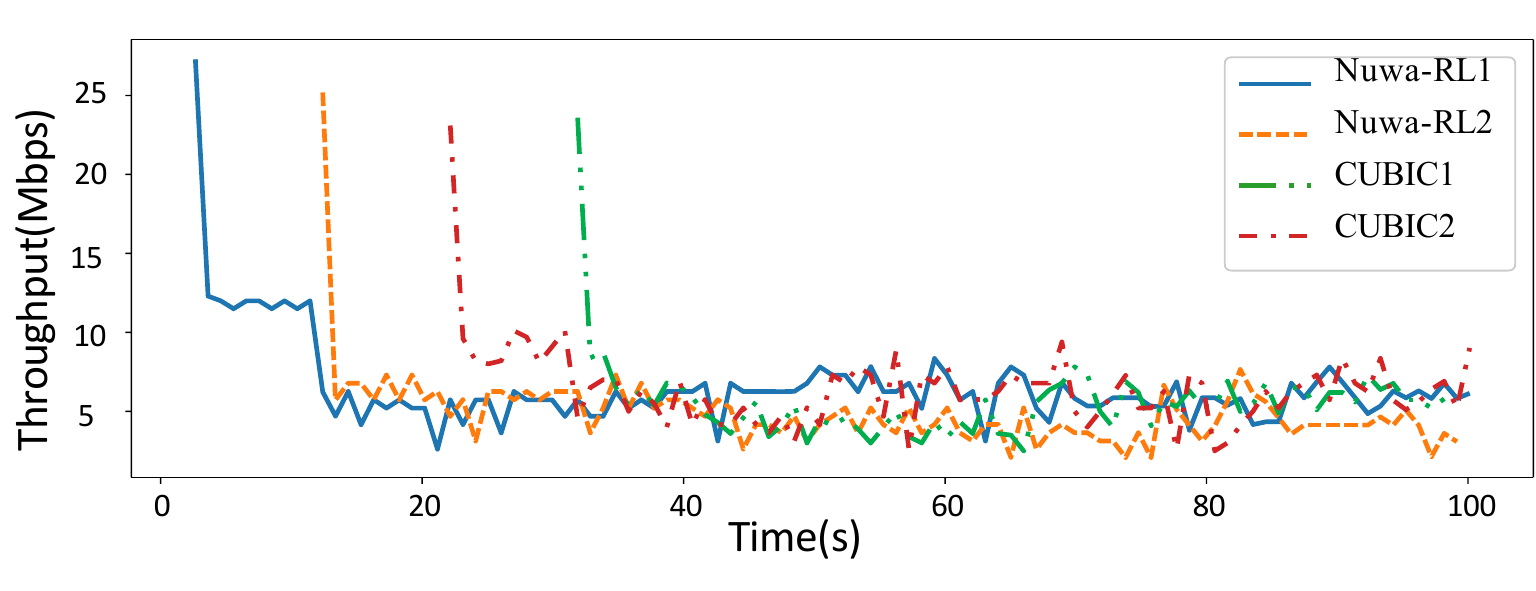}
    \caption{Fairness test of Nuwa-RL algorithm.
}
    \label{fig15}
\end{figure}

The experimental results are shown in Fig.~\ref{fig15}. After training, Nuwa-RL can effectively handle scenarios with multiple flows. When a new Nuwa-RL flow is added, the two Nuwa-RL flows can quickly share the link bandwidth fairly. When a CUBIC flow is added, Nuwa-RL can also adjust in time and share the bandwidth fairly with the CUBIC flow within 10 seconds.

\section{In-depth Analysis} \label{Sec}

\begin{figure}
    \centering
    \includegraphics[width=1.1\linewidth]{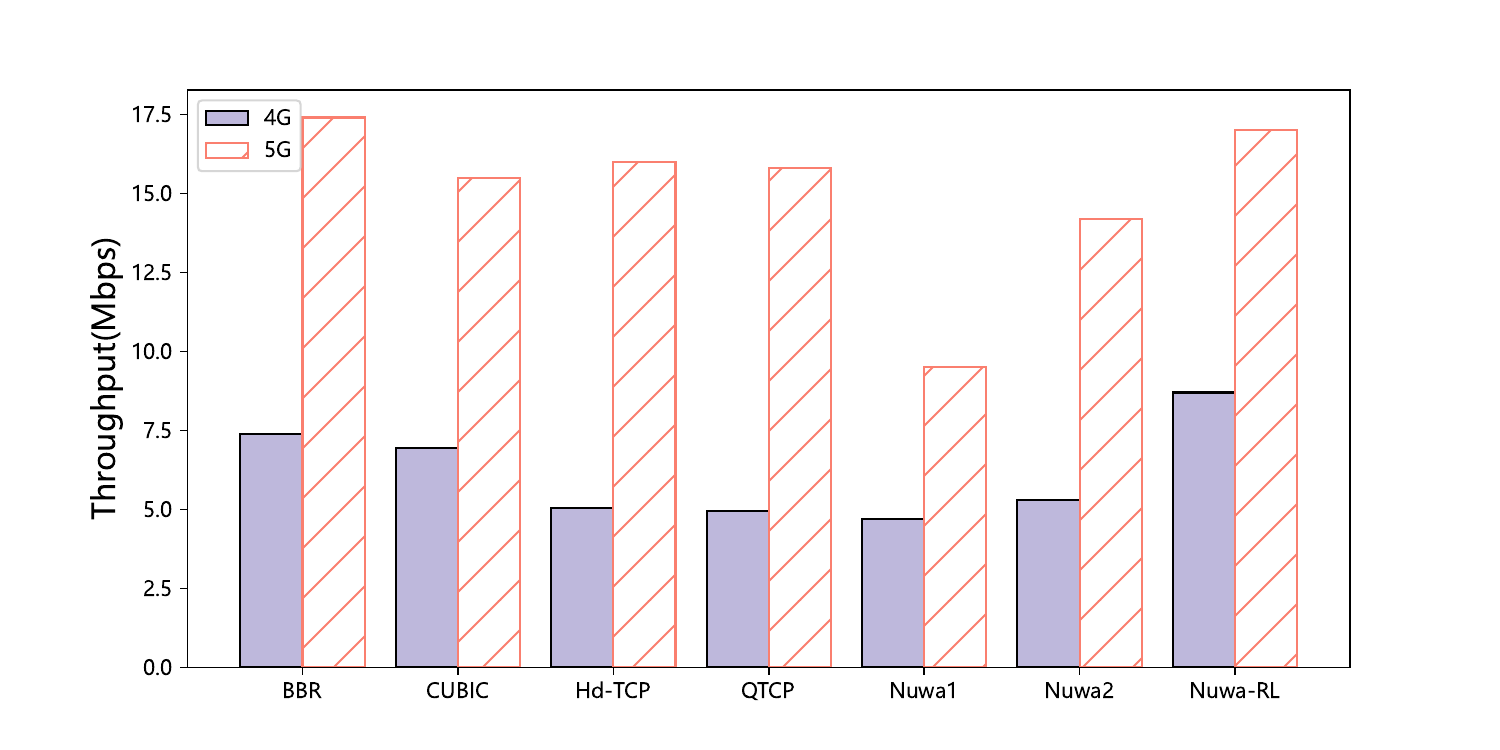}
    \caption{Throughput performance of different algorithms in 4G and 5G environments.
}
    \label{fig17}
\end{figure}

We deploy the 4G and 5G traces we collect in Cellsim and evaluate the throughput performance of our algorithm using the iperf tool. The algorithms we test include BBR, CUBIC, Hd-TCP, QTCP\cite{a37:w}, Nuwa1 ($k=2$), Nuwa2 ($k=7$), Nuwa-RL, among which Hd-TCP and QTCP are reinforcement learning algorithms.The experiment run for 80 seconds, and Fig.~\ref{fig17} shows the average throughput performance of multiple experiments. In both 4G and 5G networks, Nuwa1's throughput performance is the worst among all algorithms, while Nuwa2's throughput has improved significantly, indicating that the selection of $k$ value has a significant impact on the throughput of the Nuwa algorithm. Meanwhile, Nuwa-RL performs well in terms of throughput in both 4G and 5G networks. Therefore, our approach of using reinforcement learning to adjust the parameter $k$ can assist Nuwa-RL in maintaining good throughput performance in wireless networks. Furthermore, we also observe that in a 5G network environment, the throughput of Nuwa-RL is slightly lower than that of BBR. We attribute this phenomenon to the following reasons: 1) The BBR algorithm can generally maintain good throughput in most cases, but performance degradation might occur during significant network variations. The 5G traces we collected might not exhibit sufficiently dramatic fluctuations. In reality, 5G networks could experience more pronounced variability; 2) The Nuwa algorithm and Nuwa-RL algorithm are based on the CUBIC algorithm, so there are certain limitations. Overall, setting a reasonable value of $k$ according to the network environment can optimize the performance of the Nuwa algorithm.

\begin{figure*}[!t]
\centering
\subfigure[Shanghai 10M]{
\includegraphics[width=6cm]{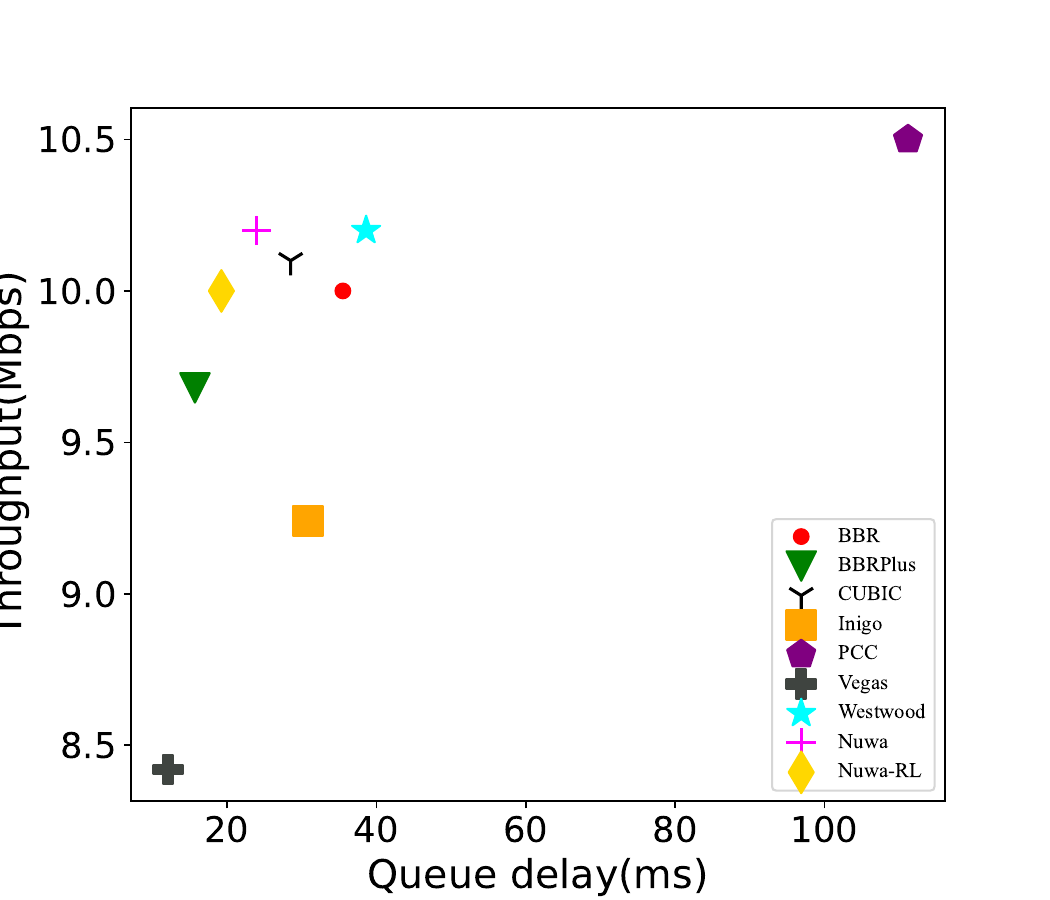}
}%
\subfigure[Beijing 10M]{
\includegraphics[width=6cm]{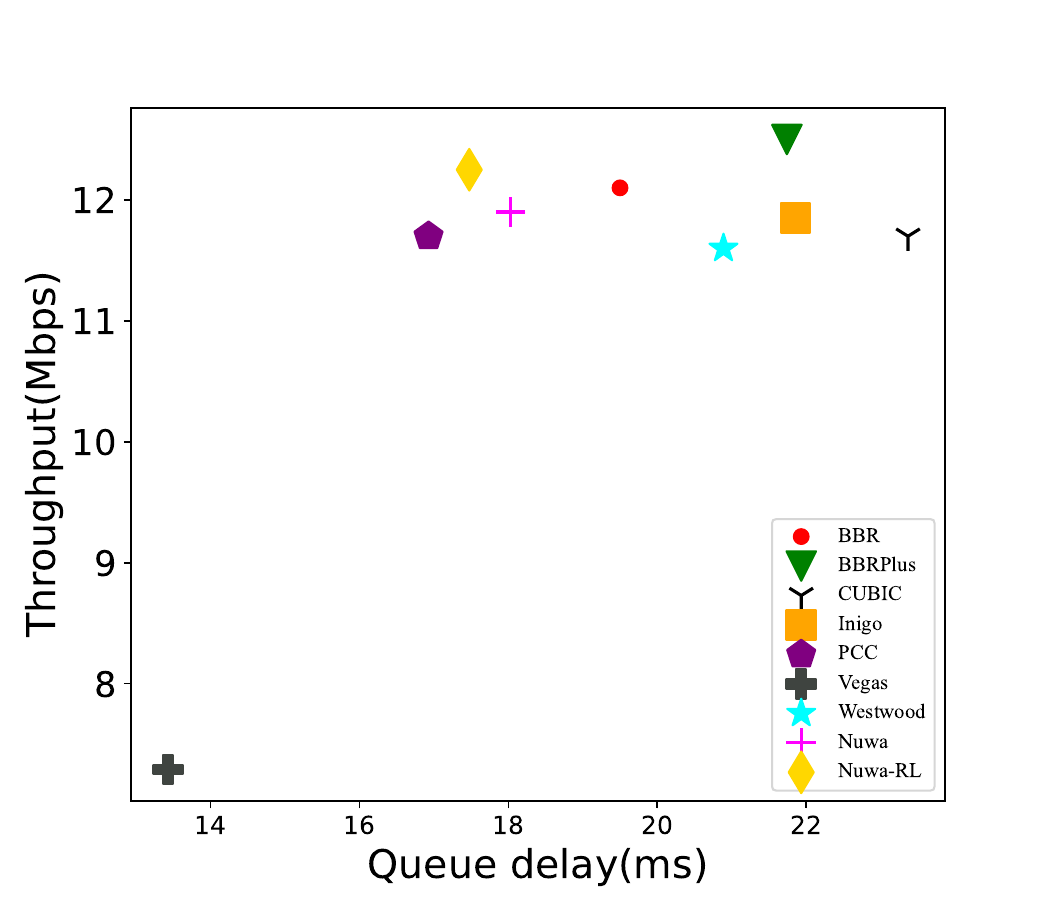}
}%
\subfigure[Guiyang 10M]{
\includegraphics[width=6cm]{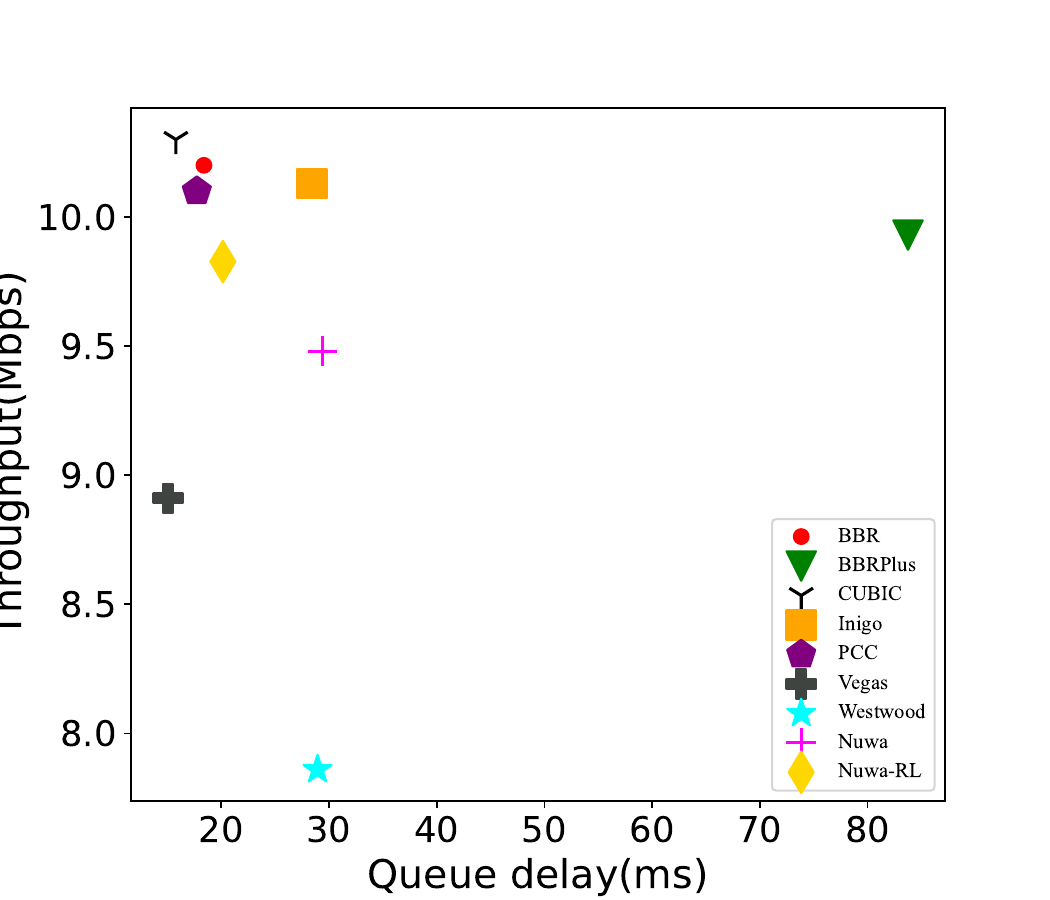}
}%
\vspace{-2.5mm}

\subfigure[Shanghai 20M]{
\includegraphics[width=6cm]{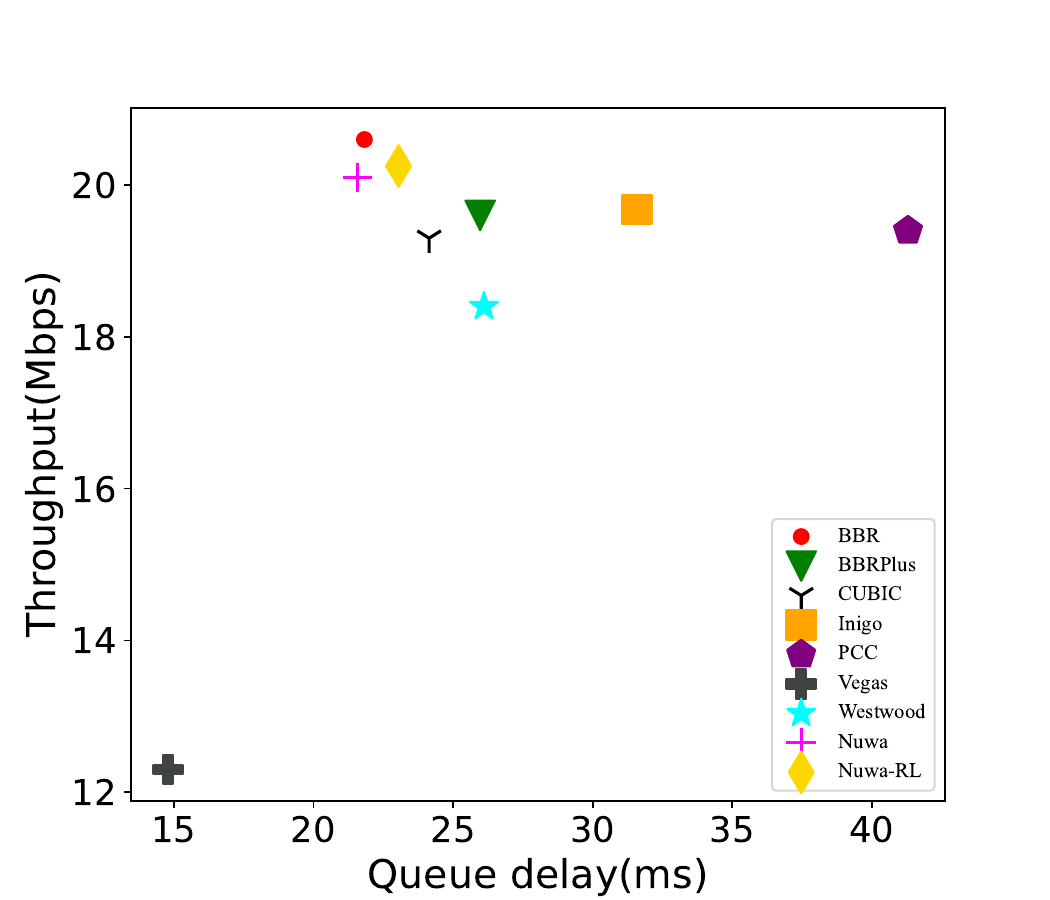}
}%
\subfigure[Beijing 20M]{
\includegraphics[width=6cm]{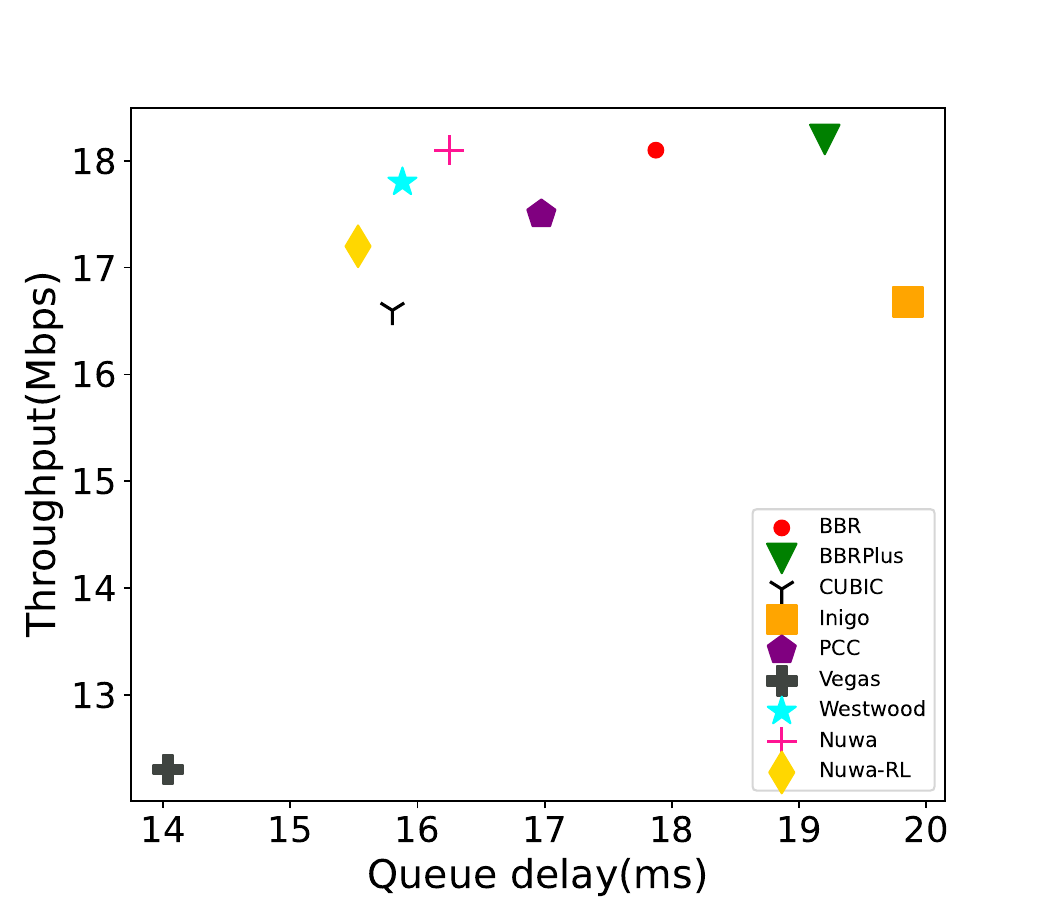}
}%
\subfigure[Guiyang 20M]{
\includegraphics[width=6cm]{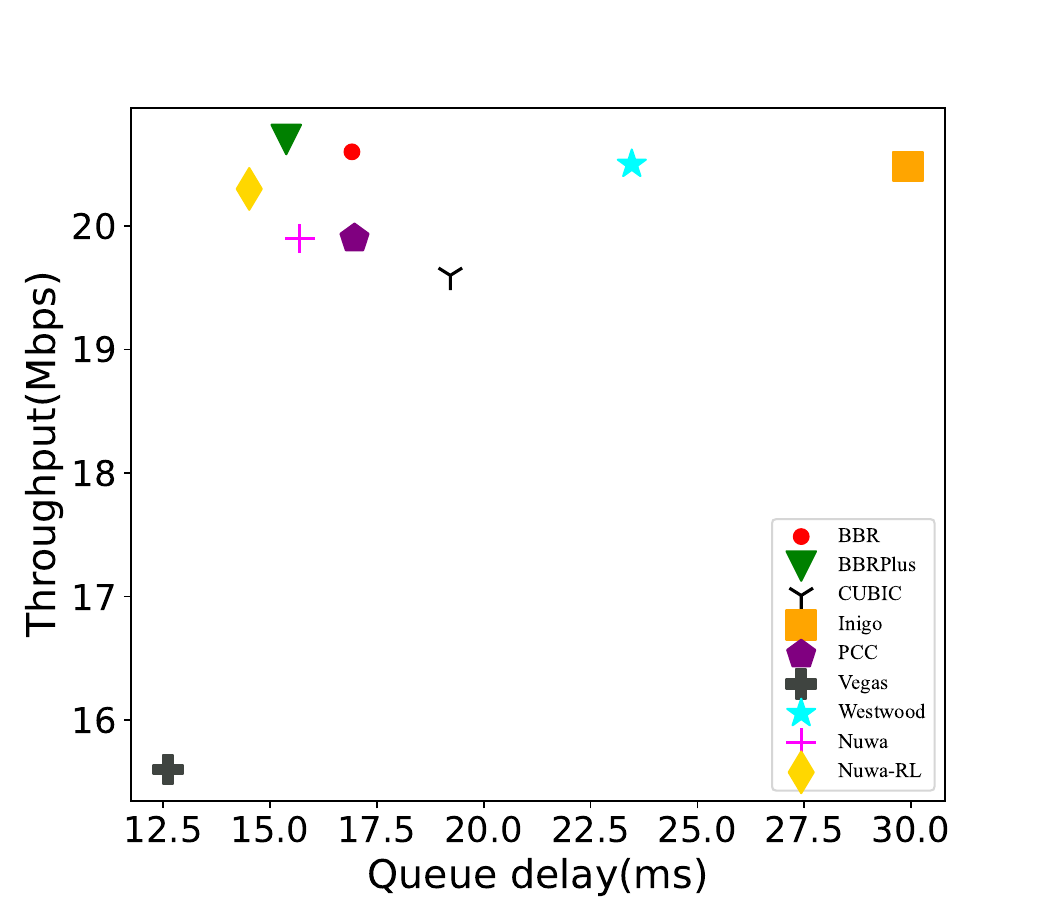}
}%
\vspace{-2.5mm}

\subfigure[Shanghai 50M]{
\includegraphics[width=6cm]{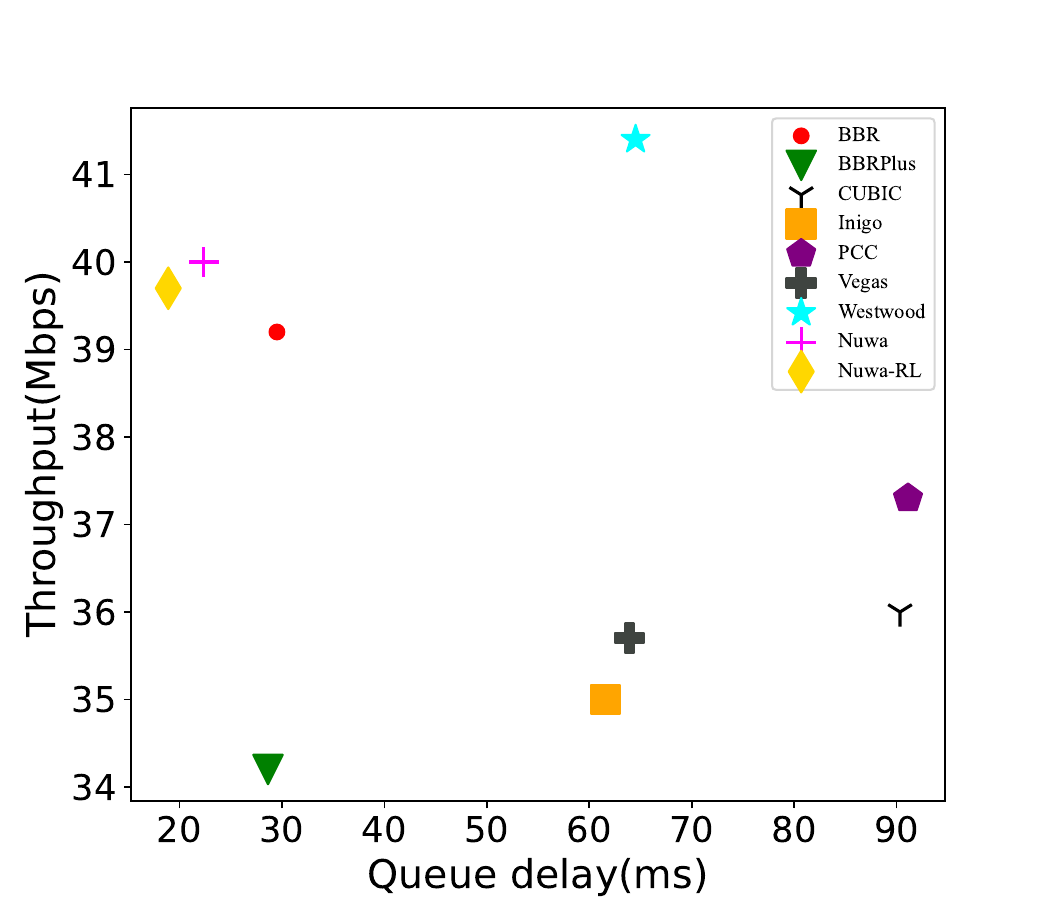}
}%
\subfigure[Beijing 50M]{
\includegraphics[width=6cm]{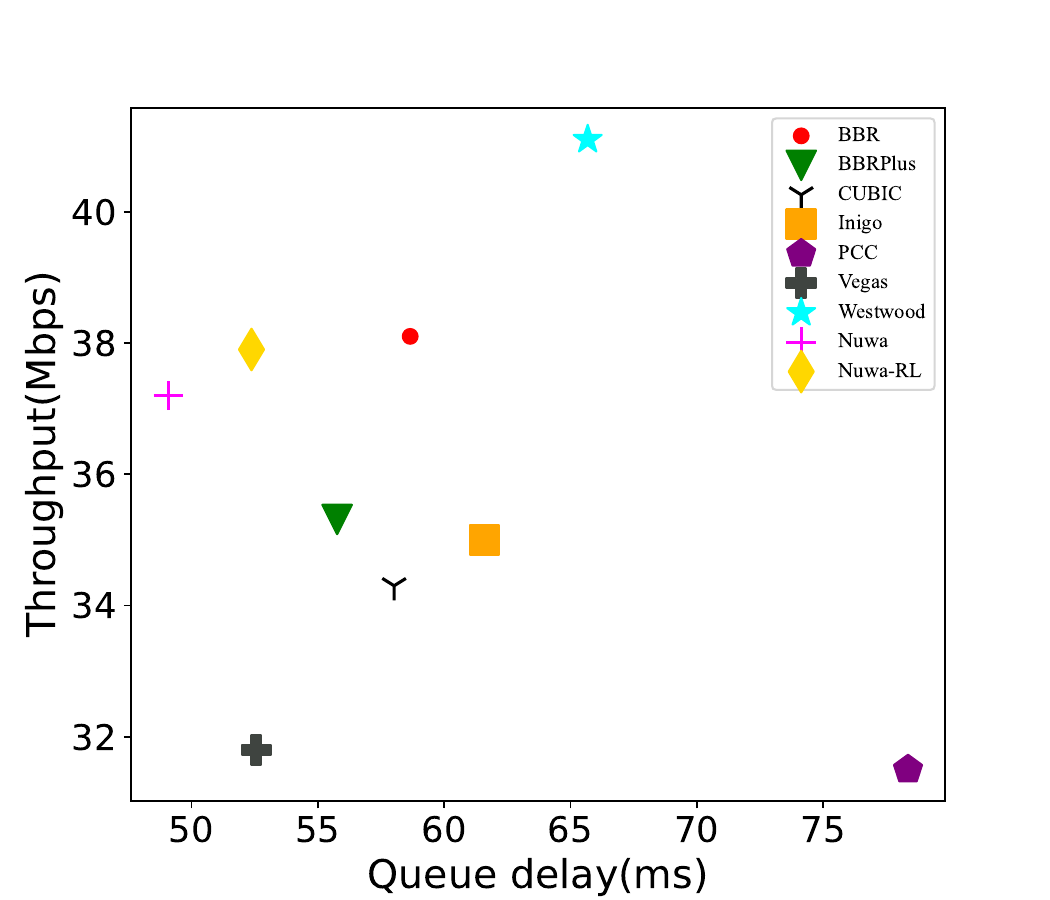}
}%
\subfigure[Guiyang 50M]{
\includegraphics[width=6cm]{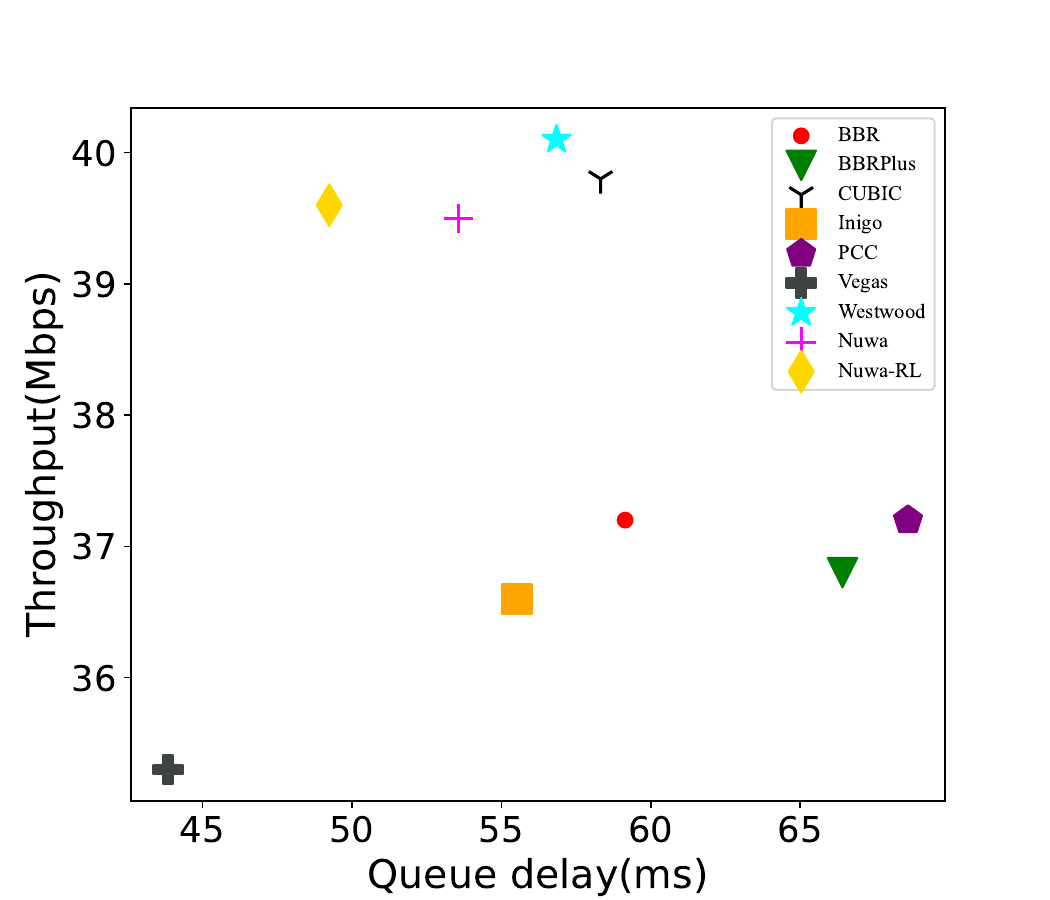}
}%
\caption{Comparison of the throughput and latency of the proposed algorithms with the baseline algorithms in different network environments.}
\label{fig16}
\end{figure*}

 We set up different 5G network scenarios and further evaluate the Nuwa algorithm ($k=7$) and Nuwa-RL algorithm from the perspectives of throughput and latency, and compare them with baseline algorithms. The experimental topology is shown in Fig.~\ref{fig1}. We use Nginx\cite{a36:L} to establish a web service on a cloud server and use it to provide file download services. To comprehensively evaluate the experiments, we deploy servers in different regions, including Shanghai, Beijing, and Guiyang, covering short, medium, and long distances to test the throughput and queue delay control of the algorithms under different distances. For the same region, we also set up three different server bandwidths, 10M, 20M, and 50M, to test whether the algorithms can maintain good performance under different bandwidths. We submit a file download request in Ningbo, and the server in the corresponding city provides the service. The experimental results are shown in Fig.~\ref{fig16}, which displays the average results of multiple experiments for each algorithm.

 From the figure, it can be seen that congestion control algorithms exhibit performance differences in different network scenarios. This is mainly because most congestion control algorithms are designed for specific network scenarios, which enables them to achieve efficient transmission in specific scenarios but also loses their adaptability to the environment. As the network bandwidth increases, the performance differences between algorithms will become more significant. The increase in transmission distance further challenges the algorithms, as the farther the distance, the greater the likelihood of queuing and packet loss during transmission. We can also see that delay-based algorithms do have stronger delay control capabilities than loss-based algorithms. For example, the Vegas algorithm, which is completely based on delay, achieves the minimum transmission delay in almost all test scenarios, but at the cost of significant throughput loss. Inigo uses independent delay-based algorithms at sender-side and receiver-side, but also sacrifices some throughput, especially in high-bandwidth scenarios. Meanwhile, loss-based algorithms such as CUBIC, which constantly fill the buffer, can achieve good throughput, especially in long-distance transmission scenarios, but their delay control capability is very unstable and shows significant differences in different network scenarios. Although the Westwood algorithm did not perform outstandingly, it can make a trade-off between throughput and delay, and its performance is relatively stable in different scenarios. BBR, BBRPlus, and PCC are all probing-based congestion control algorithms, but in the case of high bandwidth, the performance of PCC is far inferior to BBR and BBRPlus.

 Nuwa and Nuwa-RL both maintain good performance in all testing scenarios. This is because the congestion control implemented at receiver-side is based on real network feedback, and the one-way queuing delay can better reflect congestion conditions in the network. Nuwa and Nuwa-RL can adjust the window more reasonably, thus avoiding algorithm performance degradation. In addition, compared to Nuwa, Nuwa-RL exhibits greater environmental adaptability. This is due to reinforcement learning's ability to learn from the network environment and adjust Nuwa's key parameter $k$ in a timely manner, making Nuwa's window adjustment more in line with network requirements. It has been proven that accurate estimation of queuing delay can improve algorithmic transmission performance, while reinforcement learning can enhance the algorithm's environmental adaptability, making it applicable in a wider range of scenarios.

\section{Conclusion}
In this paper, we propose a new TCP framework called Nuwa, which is a receiver-driven congestion control framework. Nuwa implements the congestion avoidance phase at receiver-side, which uses one-way queue delay to observe the network congestion status, and receiver-side estimates the available window value for data transmission based on the actual feedback of the network. By setting specific target delays for different applications, the ability to occupy bandwidth of Nuwa is improved, avoiding the problem of throughput degradation when competing with loss-based algorithms. It has been experimentally proven that by configuring parameters reasonably, Nuwa can effectively control queue delays and maintain good throughput in wireless networks. We also point out that utilizing reinforcement learning to enhance Nuwa's environmental adaptability can further improve Nuwa's performance in a wider range of wireless network scenarios. In addition, this framework also provides a possibility to solve the problem of inaccurate bandwidth prediction for DASH caused by loose coupling with TCP. Based on Nuwa, in the future, emphasis can be placed on researching how to transmit the underlying information of the receiver-side to DASH, and how DASH utilizes this information to enhance the accuracy of its bandwidth prediction.

 \section*{Acknowledgment}

This work was supported in part by National Natural Science Foundation of China (61601252), Natural Science Foundation of Zhejiang Province (LY20F020008).

\vfill

\end{document}